\documentclass[11pt]{article}

\usepackage{color}
\usepackage{graphicx}
\usepackage{amsmath}
\usepackage{amssymb}
\usepackage{xspace}
\usepackage[small]{subfigure}
\usepackage[numbers,compress]{natbib}
\usepackage[hyperfootnotes=false]{hyperref}
\usepackage{tcolorbox}

\newlength{\fighskip} \fighskip=2pt
\newlength{\figvskip} \figvskip=3pt

\usepackage{mciteplus} 
\usepackage{dcolumn}
\usepackage{bm}
\usepackage{verbatim}
\usepackage{amscd}
\usepackage{amsfonts}
\usepackage{setspace}
\usepackage{amsthm}
\usepackage{enumerate}
\usepackage{mathtools}

\begin{document}

\title{\bf 
Soft thermodynamics of gravitational shock wave
}
\author{
Shuwei Liu and Beni Yoshida\\ 
{\em \small Perimeter Institute for Theoretical Physics, Waterloo, Ontario N2L 2Y5, Canada} }
\date{}

\maketitle

\begin{abstract}
The gravitational shock waves have provided crucial insights into entanglement structures of black holes in the AdS/CFT correspondence. 
Recent progress on the soft hair physics suggests that these developments from holography may also be applicable to geometries beyond negatively curved spacetime. 
In this work, we derive a remarkably simple thermodynamic relation which relates the gravitational shock wave to a microscopic area deformation.
Our treatment is based on the covariant phase space formalism and is applicable to any Killing horizon in generic static spacetime which is governed by arbitrary covariant theory of gravity. 
The central idea is to probe the gravitational shock wave, which shifts the horizon in the $u$ direction, by the Noether charge constructed from a vector field which shifts the horizon in the $v$ direction. 
As an application, we illustrate its use for the Gauss-Bonnet gravity. 
We also derive a simplified form of the gravitational scattering unitary matrix and show that its leading-order contribution is nothing but the exponential of the horizon area: $\mathcal{U}=\exp(i \text{Area})$.

\end{abstract}

\tableofcontents

\section{Introduction}

Recent developments at the interface between quantum gravity and quantum information theory in the AdS/CFT correspondence have provided useful tools to address conceptual puzzles concerning quantum aspects of black holes on less ambiguous settings~\cite{Page:1993aa, Hayden07, Verlinde12, Almheiri:2015ac, Pastawski15b, Hosur:2015ylk, Dong2016, Penington19, Almheiri19}. At the heart of these progresses is the improved understanding on how the structure of quantum entanglement changes dynamically under the gravitational backreaction from the infalling and outgoing matter~\cite{Hooft:1987aa, tHooft:1987vrq, Verlinde:1991iu, tHooft:1990fkf, Jacobson:1993hn, Schoutens:1994st, Gubser:2008pc, Giddings:2004xy, Shenker:2013pqa, Kitaev_unpublished, Roberts:2014isa, Shenker:2014cwa, Kitaev:2017awl, Yoshida:2017aa, Roberts:2018aa, Beni18, Beni19, Yoshida:2021aa}. The gravitational shock wave geometries~\cite{Aichelburg:1970dh, Dray:1984ha, Dray:1985yt, Hotta:1993aa, Sfetsos:1994xa, Cai:1999aa, Campanelli:1995ex, Alishahiha:2016cjk, BenTov:2017kyf} provide a particularly useful family of analytically tractable models of the gravitational backreaction that can be induced by infalling massless particles, or equivalently, perturbations on the boundary by light operators. 

Despite extensive studies in the past decades and recent revivals in the AdS/CFT correspondence, the quantum origin of the gravitational shock waves still remains elusive. The gravitational shock waves are often interpreted as low-energy excitations which constitute the microscopic degrees of freedom of the quantum black hole. It is, however, unclear how the gravitational shock waves are ever quantized and account for the finite value of Bekenstein-Hawking entropy in a concrete manner. Also, if one hopes to study geometries beyond the AdS/CFT correspondence (\emph{e.g.} asymptotically flat spaces or systems without the spatial infinity as in the de Sitter space), additional technical and conceptual subtleties often hinder naive application of holographic intuitions~\footnote{For instance, the absence of the spatial infinity may suggest that the structure of quantum entanglement in the de Sitter space differs crucially from that of the AdS space or the asymptotically flat space.}
~\footnote{
See~\cite{Geng:2020aa, Geng:2021aa} for other studies of entanglement structure in the de Sitter space. 
}. Hence, as a first step, it is desirable to develop a universal framework to characterize the gravitational shock waves through some microscopic lens in a way applicable to geometries beyond the AdS/CFT correspondence as well.

Recent rapid progress on the soft hair of black holes in an asymptotically flat space provide useful hints toward such a goal with successful derivations of the Bekenstein-Hawking entropy~\cite{Castro:2010fd,Carlip:2017xne,Haco:2018ske,Aggarwal:2019iay,Averin:2019zsi,Chen:2020nyh}
~\footnote{See~\cite{Bondi:1962px, Sachs:1962zza, Carlip:1998wz,Carlip:1999cy,Koga:2001vq,Donnay:2015abr, Donnelly:2016auv,Donnay:2016ejv,Speranza:2017gxd,Compere:2018aar} for samples of earlier works and relevant developments.}. 
It has been also pointed out that the linearized gravitational shockwaves can be realized as BMS supertranslations via certain choices of parameters~\cite{Strominger:2014pwa,Hawking:2016msc,Hawking:2016sgy}. Furthermore, some general expressions of thermodynamic relations for soft charges have been derived for the Schwarzschild black hole and several other geometries~\cite{Donnay:2018aa}. 

While these developments may provide a further insight into quantum aspects of the gravitational shock waves, it remains unclear how the soft charges may be measured in a concrete setting. In addition, the derived thermodynamic relations, associated with soft charges, are in rather abstract forms without concrete physical interpretations. As such, the implications of the soft hair physics on conceptual puzzles of quantum black holes still remain vague, in comparison with concrete developments within the framework of the AdS/CFT correspondence. What is currently missing is an effort to tie these two developments. Hence, deriving a microscopic thermodynamic characterization of the gravitational shock waves on a unified footing may be the necessary first step.

In this paper, we derive the microscopic thermodynamic relations which are localized on the near horizon region and are sensitive to the gravitational shock wave in a concrete form. Our treatment is applicable to arbitrary gravitational shock waves at a bifurcating horizon in generic static spacetime (including AdS, dS, and asymptotically flat spaces), and is valid for arbitrary covariant theory of gravity which may be beyond general relativity (\emph{e.g.} higher-derivative gravity theories). Furthermore, our thermodynamic relation provides a concrete physical interpretation of the gravitational shock waves, or soft charges. In particular, we derive a remarkably simple formula which relates the horizon area deformation induced by two intersecting gravitational shock waves to the incoming and outgoing energy sources. We will then find that the soft charge associated with the gravitational shock waves is nothing but the horizon area in the presence of two intersecting gravitational shock waves.

Our main technical machinery is the covariant phase space formalism, namely Wald's Noether charge method. It turns out that the original treatment by Wald, which focuses on contributions at the linear order of field variations, is not sufficient to probe the gravitational shock wave. For this reason, we will extend the Noether charge method to include variations of the matter fields at \emph{arbitrary order}. This refinement enables us to obtain the microscopic thermodynamic relations (as in Eq.~\eqref{eq:summary}) which directly associate the area deformation to the energy momentum tensor variation. One benefit of the covariant phase space formalism is that it is applicable to arbitrary covariant theory of gravity. As an application, we demonstrate that, in the Gauss-Bonnet gravity~\cite{lovelock:1971yv,Stelle:1976gc, Brown:1980qq, Adler:1982ri}, the gravitational shock wave equations indeed follow from the microscopic thermodynamic relation with much simplified calculations.

Our result also suggests a concrete framework to physically characterize (or, even measure) soft charges via out-of-time order correlation (OTOC) functions. The leading-order (in $1/G_{N}$) behaviours of OTOCs are dominated by the gravitational scattering unitary matrix which typically takes the following form; $\mathcal{U} = \exp \left[ i \iint  d \Omega_1 d \Omega_2 P_{\text{out}}(\Omega_1) f(\Omega_1,\Omega_2) P_{\text{in}}(\Omega_2)   \right]$. Here $f(\Omega_1,\Omega_2)$ is the green function for the shock wave equation, and $P_{\text{in}}$ and $P_{\text{out}}$ account for the incoming and outgoing energy fluxes. Our thermodynamic characterization enables us to rewrite the above scattering matrix into a remarkably simple form:
\begin{align}
\mathcal{U} = \exp \big( i \text{Area} \big)
\end{align}
up to some multiplicative factor in the phase. Here ``$\text{Area}$'' represents the horizon area in the presence of two intersecting gravitational shockwaves due to $P_{\text{in}}$ and $P_{\text{out}}$. Thus, the phase factor of the scattering matrix is proportional to the horizon area (or equivalently, the soft charge of an infalling matter measured by an outgoing matter).

\subsection{Main result}

Here we present a summary of our main result. We study a bifurcating Killing horizon in generic static spacetime of the following form:
\begin{equation}
\begin{split}
ds^2 = 2 F(u,v) dudv + G(u,v) h_{ij} dx^i dx^j, \qquad 
 T = 2T_{uv}dudv + T_{uu}dudu + T_{vv}dvdv  + T_{ij}dx^{i}dx^{j} \label{eq:summary_original}
\end{split}
\end{equation}
where $u=0$ and $v=0$ corresponds to the bifurcating horizons. We assume that the timelike vector $\partial_t = \kappa( -u \partial_u + v \partial_v)$ is a Killing vector of the spacetime. The gravitational shock wave can be generated by shifting this spacetime as
\begin{align}
\tilde{v}=v + \theta(u)\alpha(x)
\end{align}
where $\theta(u)$ is a step function and $x$ collectively denotes angular profiles. At the leading order, the metric variation can be computed by considering the horizon translation vector $\hat{\alpha}$:
\begin{align}
\delta g_{\mu\nu} =\theta(u) \mathcal{L}_{\hat{\alpha}} g_{\mu\nu}, \qquad \hat{\alpha} \equiv \alpha(x)\partial_v
\end{align} 
where $\mathcal{L}_{\hat{\alpha}}$ is the Lie derivative with respect to $\hat{\alpha}$. To satisfy the Einstein equation, an additional energy momentum tensor needs to be introduced:
\begin{align}
{T^{(p)}}^{vv} = T^{(p)}(x)\delta(u).
\end{align}
For general relativity (\emph{i.e.} $L^{(G)}=\frac{1}{16\pi}  (R-2\Lambda)$), Sfetsos derived the shock wave equation of motion which relates the shift profile $\alpha(x)$ and the energy momentum profile $T^{(p)}(x)$ via a certain differential equation~\cite{Sfetsos:1994xa}. In this paper, we will work on arbitrary covariant theory of gravity (\emph{i.e.} arbitrary covariant Lagrangian $L^{(G)}$) which may be beyond general relativity.

We will investigate the shock wave geometry by using the covariant phase space formalism~\cite{doi:10.1063/1.528839, Crnkovic:1986ex, Lee:1990nz, Wald:1993nt, Iyer:1994ys, Wald:1999wa}~\footnote{See \cite{Khavkine:2014kya} for a review.} and derive a microscopic thermodynamic formula which captures low-energy excitations on the Killing horizon. The standard method utilizes the timelike Killing vector $\xi^{(t)} = \partial_t$ to construct the Noether charge $\mathbf{Q}_{\xi}$ and relate the horizon area (at $u=v=0$) to the ADM thermodynamic parameters at asymptotic infinity. It turns out, however, that the area variation due to the gravitational shock wave vanishes at $u\rightarrow 0$, and hence non-trivial thermodynamic relations cannot be derived. 

In this paper, we study the gravitational shock wave by using the Noether charge $\mathbf{Q}_{\hat{\beta}}$ constructed from another horizon translation vector $\hat{\beta}$ in the $u$ direction:
\begin{align}
\hat{\beta} = \beta(x) \partial_u.
\end{align}
In particular, we probe the gravitational shock wave, which was generated by the shift $v \rightarrow v + \alpha(x)$ in the $v$ direction, by another shift $u \rightarrow u + \beta(x)$ in the $u$ direction. For general relativity, this enables us to derive the following microscopic thermodynamic relation:
\begin{align}
\text{Area}(\beta(x),\alpha(x)) - \text{Area}(0,0) = \int \sqrt{-g}\beta(x) T^{(p)}(x) dx \label{eq:summary}
\end{align}
up to a multiplicative factor. Here $\text{Area}(\beta(x),\alpha(x))$ corresponds to the total area of the sphere at  $u=\beta(x)$ and $v=\alpha(x)$. For a generic covariant Lagrangian $\mathbf{L}^{(G)}$, the area term would correspond to the Wald's geometric entropy. This thermodynamic relation can be rewritten as 
\begin{align}
E_{\text{shock}} \beta(x) = 0
\end{align}
such that $E_{\text{shock}}=0$ corresponds to the shock wave equation of motion for $\alpha(x)$ and $T^{(p)}(x)$. 

We will also study the gravitational scattering unitary matrix:
\begin{align}
\mathcal{U} = \exp \left[ i \iint  d \Omega_1 d \Omega_2 P_{\text{out}}(\Omega_1) f(\Omega_1,\Omega_2) P_{\text{in}}(\Omega_2)   \right]. 
\end{align}
Let us assume that $P_{\text{out}}$ and $P_{\text{in}}$ induce the shifts $\alpha(x)$ and $\beta(x)$. Then, for general relativity, the above thermodynamic characterization enables us to rewrite $\mathcal{U}$ as follows:
\begin{align}
\mathcal{U} = \exp \Big( i  (\text{Area}(\alpha(x),\beta(x))  -\text{Area}(0,0)  )  \Big)
\end{align}
up to a multiplicative factor for the area. Hence, the scattering matrix is the exponential of the horizon area (or equivalently, the soft charge).

\subsection{Relation to previous works}

The microscopic relation between the area deformation and the gravitational shock wave for the Schwarzschild black hole, for the case with $T^{\mu\nu}=0$ (\emph{i.e.} a ``sourceless'' shock wave), was initially communicated to us from Yoni BenTov~\footnote{BenTov told us that he learned it from Alexei Kitaev.}. Our contribution is to include the energy momentum tensor and extend this observation to a bifurcate Killing horizon in generic static spacetime for arbitrary covariant theory of gravity. The covariant phase space formalism has been recently applied to the Jackiw-Teitelboim (JT) gravity which is a tractable model of the gravitational backreaction~\cite{Harlow:2020aa}. 

In an asymptotically flat space, Hawking, Perry and Strominger have identified a concrete way of generating the gravitational shock wave from BMS supertranslation and superrotation on the boundary of the asymptotic flat spacetime~\cite{Hawking:2016sgy}. Namely they provided a generic formula to prove the shock wave can be created by a supertranslation, by using the Noether charge constructed from a supertoration. This result was further investigated and extended, see~\cite{Donnay:2018aa, Ward_thesis, Pasterski20, Kirklin:2018aa} for instance. It is worth clarifying our contributions and novelties in comparison with these previous works. First, previous works study the linearized shock wave near the horizon whereas our treatment applies to the shock wave which runs exactly on the horizon as well. Second, our thermodynamic relation is cast in a concrete form, and hence its physical interpretation as a microscopic Hilbert space of a black hole becomes transparent. Third, as already noted above, our treatment applies to any bifurcate Killing horizon in generic static spacetime for arbitrary covariant theory of gravity.

It should be noted that, while our primary motivation stems from quantum aspects of black holes, the treatments in the present paper are entirely classical. Also, our treatment is localized near the horizon and does not consider asymptotic infinity or boundaries. As such, we do not discuss BMS transformations in a direct manner. 

\subsection{Organization}

The paper is organized as follows. In section~\ref{sec:Noether}, we present a review of Wald's Noether charge method. In section~\ref{sec:refinement}, we will make certain refinements to the Wald's entropy formula so that it is valid up to higher order of matter field variations. In section~\ref{sec:shockwave}, we present a review of gravitational shock wave on a bifurcate Killing horizon in generic static spacetime. In section~\ref{sec:Killing}, we investigate the variations of the metric due to horizon translations. In section~\ref{sec:thermo}, we derive a microscopic thermodynamic relation on the horizon. In section~\ref{sec:area}, we demonstrate that the shockwave equations of motion indeed follows from the area deformation. In section~\ref{sec:application}, we apply our formalism to the Gauss-Bonnet gravity. In section~\ref{sec:OTOC}, we briefly discuss the gravitational scattering matrix. In section~\ref{sec:outlook}, we conclude with discussions.


\section{Review of the Noether charge method}
\label{sec:Noether}

In this section, we present a self-contained review of the Noether charge method in the presence of matter fields. Readers who are familiar with the method may skip to the next section.

\subsection{Variation of Lagrangian}
\label{subsec:General form}

Bold letters are used to represent differential forms. Consider the following diffeomorphism covariant Lagrangian $d$-form on a $d$-dimensional manifold~\footnote{
The diffeomorphism covariance condition is  
$\mathcal{L}_{\xi}\mathbf{L}(\psi)=\frac{\partial \mathbf{L}}{\partial \psi}\mathcal{L}_{\xi}\psi$.
}:
\begin{align}
\mathbf{L} = \mathbf{L}^{(G)} + \mathbf{L}^{(M)}  \label{eq:separation}
\end{align}
where $\mathbf{L}^{(G)}$ is the gravitational Lagrangian $d$-form and $\mathbf{L}^{(M)}$ is the matter Lagrangian $d$-form. 
Here we assumed that the Lagrangian can be separated into two parts $\mathbf{L}^{(G)}$ and $\mathbf{L}^{(M)}$~\footnote{More precisely, we assume that $\mathbf{L}^{(G)}$ consists of terms which do not involve matter fields $\phi$ and their covariant derivatives $\nabla_{\mu_1} \ldots \nabla_{\mu_m} \phi$ whereas $\mathbf{L}^{(M)}$ consists of terms which do not involve  Riemann tensors $R_{abcd}$ and their covariant derivatives $\nabla_{\mu_1} \ldots \nabla_{\mu_m} R_{abcd}$. As such, Riemann tensors $R_{abcd}$ do not directly couple to matter fields (e.g. terms like $R \phi^2$ do not appear) in the Lagrangian. Such matter fields are often said to be minimally coupled to gravity. Gravitational shock waves with non-minimal couplings have been studied in~\cite{Superconvergence}.
}. 
At the linear order in $\delta g_{\mu\nu}, \delta \phi$, the variations can be expressed as
\begin{equation}
\begin{split}
\delta \mathbf{L}^{(G)} &= {\mathbf{E}^{(G)}}^{\mu\nu} \delta g_{\mu\nu} + d \mathbf{\Theta}^{(G)} (g_{\mu\nu},\delta g_{\mu\nu}), \\
\delta \mathbf{L}^{(M)} &= \frac{1}{2}  \pmb{\epsilon} T^{\mu\nu} \delta g_{\mu\nu} + \mathbf{E}^{(\phi)} \delta \phi  + d \mathbf{\Theta}^{(M)}(g_{\mu\nu}, \phi,\delta \phi ) \label{eq:variation}
\end{split}
\end{equation}
where $\phi$ collectively denotes all the matter fields.
Here $\pmb{\epsilon}$ denotes the volume form $\pmb{\epsilon} = \mathbf{d^d x} \sqrt{-g}$ and $\mathbf{\Theta}^{(G)}, \mathbf{\Theta}^{(M)}$ are called symplectic potentials
~\footnote{
Since the matter Lagrangian does not involve $R_{abcd}$ and its covariant derivatives, the matter symplectic potential $\mathbf{\Theta}^{(M)}$ can be constructed so that it does not depend on $\delta g_{\mu\nu}$. See Section 3 of Iyer and Wald for details~\cite{Iyer:1994ys}.}. 
These variations lead to equations of motion:
\begin{align}
{\mathbf{E}^{(G)}}^{\mu\nu} + \frac{1}{2} \pmb{\epsilon}  T^{\mu\nu} =0, \qquad \mathbf{E}^{(\phi)}=0 \label{eq:motion}
\end{align}
where the energy momentum tensor is defined by~\footnote{
The Euler-Lagrange derivative is defined by
\begin{align}
\frac{\delta L}{\delta \phi} = \frac{\partial L}{\partial \phi} - \partial_{\mu} \frac{\partial L}{\partial \partial_{\mu} \phi} + \partial_{\mu} \partial_{\nu} \frac{\partial L}{\partial \partial_{\mu}\partial_{\nu} \phi} + \cdots. \notag
\end{align}
}
\begin{align}
T^{\mu\nu} \equiv \frac{2}{\sqrt{-g}} \frac{\delta (\sqrt{-g} L^{(M)})}{ \delta g_{\mu\nu} }, \qquad \mathbf{L}^{(M)}  = \pmb{\epsilon}L^{(M)}.
\end{align}
Field configurations $\{ g_{\mu\nu}, \phi \}$ which satisfy equations of motion Eq.~\eqref{eq:motion} are called on shell. 

Now consider variations under an infinitesimal diffeomorphism by an arbitrary vector field $\xi$. According to Eq.~\eqref{eq:variation}, the total on shell variation can be expressed by
\begin{align}
\hat{\delta}_{\xi} \mathbf{L} = \hat{\delta}_{\xi} \mathbf{L}^{(G)} + \hat{\delta}_{\xi} \mathbf{L}^{(M)} = d \mathbf{\Theta}_{\xi}^{(G)} + d \mathbf{\Theta}_{\xi}^{(M)} \label{eq:diffeo} \qquad \text{$\{g_{\mu\nu}, \phi\} $ on shell}
\end{align}
where $\mathbf{\Theta}_{\xi}^{(G)}\equiv \mathbf{\Theta}^{(G)}(g_{\mu\nu},\hat{\delta}_{\xi}g_{\mu\nu})$ and $\mathbf{\Theta}_{\xi}^{(M)}\equiv \mathbf{\Theta}^{(M)}(g_{\mu\nu},\phi,\hat{\delta}_{\xi}\phi)$ with $\hat{\delta}_{\xi}g_{\mu\nu}\equiv\mathcal{L}_{\xi} g_{\mu\nu}$ and $\hat{\delta}_{\xi}\phi\equiv\mathcal{L}_{\xi} \phi$. Here $\mathcal{L}_{\xi}$ denotes the Lie derivative with respect to $\xi$. It is worth emphasizing that the variation $\hat{\delta}_{\xi} \mathbf{L}$ is induced by a diffeomorphism $\hat{\delta}_{\xi}$ acting on each field $g_{\mu\nu}, \phi$.

The variation $\hat{\delta}_{\xi} \mathbf{L}$ can be expressed in another form due to the diffeomorphism covariance. The covariance of the gravity and matter Lagrangians implies
\begin{align}
\hat{\delta}_{\xi} \mathbf{L}^{(G)} = \mathcal{L}_{\xi} \mathbf{L}^{(G)}, \qquad \hat{\delta}_{\xi} \mathbf{L}^{(M)} = \mathcal{L}_{\xi} \mathbf{L}^{(M)}
\label{eq:GM-variation}
\end{align}
where $\mathcal{L}_{\xi} \mathbf{L}^{(G)}, \mathcal{L}_{\xi} \mathbf{L}^{(M)}$ are the Lie derivatives of $\mathbf{L}^{(G)}, \mathbf{L}^{(M)}$. We have
\begin{align}
\hat{\delta}_{\xi} \mathbf{L}^{(G)} = \mathcal{L}_{\xi} \mathbf{L}^{(G)} = d(\xi \cdot \mathbf{L}^{(G)}), \qquad \hat{\delta}_{\xi} \mathbf{L}^{(M)} = \mathcal{L}_{\xi} \mathbf{L}^{(M)} = d(\xi \cdot \mathbf{L}^{(M)})  \label{eq:Lie-matter}
\end{align}
where we made use of the Cartan's magic formula\footnote{Here $\xi \cdot \mathbf{\Lambda}$ indicates insertion of $\xi$ into the first argument of $\mathbf{\Lambda}$. It is also called interior product and denoted as $\iota_{\xi}\mathbf{\Lambda}$.} 
\begin{align}
\mathcal{L}_{\xi} \mathbf{\Lambda} = \xi \cdot d \mathbf{\Lambda} + d(\xi \cdot \mathbf{\Lambda}) \qquad \text{$\mathbf{\Lambda}$ : arbitrary}. \label{eq:Cartan}
\end{align}
Note that $\hat{\delta}_{\xi} \mathbf{L}^{(G)}, \hat{\delta}_{\xi} \mathbf{L}^{(M)}$ must be total derivatives according to Eq.~\eqref{eq:Lie-matter}. 

\subsection{Wald's entropy formula}

Wald introduced the following Noether current $(d-1)$-form
\begin{align}
\mathbf{J}_{\xi, \text{Wald}} \equiv \mathbf{\Theta}^{(G)}_{\xi} - \xi \cdot \mathbf{L}^{(G)}  + \mathbf{\Theta}^{(M)}_{\xi}  - \xi \cdot \mathbf{L}^{(M)}.
 \label{eq:Noether-current}
\end{align}
One can verify
\begin{align}
d \mathbf{J}_{\xi, \text{Wald}} =0 \qquad  \text{$\{g, \phi\} $ on shell}.
\end{align}
Here we used $g$ to denote the metric tensor $g_{\mu\nu}$ for brevity of notation. One can actually show that $\mathbf{J}_{\xi, \text{Wald}}$ is exact due to the fact that $\mathbf{J}_{\xi, \text{Wald}}$ is closed for all $\xi$~\cite{Lee:1990nz}. As such, the Noether charge $(d-2)$-form $\mathbf{Q}_{\xi, \text{Wald}}$ can be constructed
~\footnote{
A systematic algorithm to construct the Noether charge $\mathbf{Q}_{\xi, \text{Wald}}$ can be found in Iyer and Wald~\cite{Iyer:1994ys}.
}
\begin{align}
\exists \mathbf{Q}_{\xi, \text{Wald}} \ \ \text{such that} \ \ \mathbf{J}_{\xi, \text{Wald}} = d \mathbf{Q}_{\xi, \text{Wald}} \qquad \text{$\{g, \phi\} $ on shell}.
\end{align}

Let us now fix $\xi$ and consider the variation of the Noether current Eq.~\eqref{eq:Noether-current} at the linear orders of $\delta g_{\mu\nu}, \delta \phi$: 
\begin{equation}
\begin{split}
\delta \mathbf{J}_{\xi, \text{Wald}} &= \delta[\mathbf{\Theta}^{(G)}(\mathcal{L}_{\xi}g)] - \xi\cdot \delta \mathbf{L}^{(G)} 
+ \delta[\mathbf{\Theta}^{(M)}(\mathcal{L}_{\xi}\phi)] - \xi\cdot \delta \mathbf{L}^{(M)}
\\
&= \delta[\mathbf{\Theta}^{(G)}(\mathcal{L}_{\xi}g)] - \xi \cdot d \mathbf{\Theta}^{(G)}(\delta g)  +
\delta[\mathbf{\Theta}^{(M)}(\mathcal{L}_{\xi}\phi)] - \xi \cdot d \mathbf{\Theta}^{(M)}(\delta \phi)
\end{split}
\end{equation}
where we used the on shell conditions Eq.~\eqref{eq:motion} in the second line. Let us define the symplectic currents $\mathbf{\Omega}^{(G)}(g,\delta g, \mathcal{L}_{\xi}g )$,  $\mathbf{\Omega}^{(M)}(g,\delta g, \mathcal{L}_{\xi}g,\phi,\delta \phi, \mathcal{L}_{\xi}\phi )$ by
\begin{equation}
\begin{split}
\mathbf{\Omega}^{(G)}(g,\delta g, \mathcal{L}_{\xi}g ) &\equiv \delta[ \mathbf{\Theta}^{(G)}(g,\mathcal{L}_{\xi} g)] - \mathcal{L}_{\xi}[ \mathbf{\Theta}^{(G)}(g,\delta g)]\\
\mathbf{\Omega}^{(M)}(g,\delta g, \mathcal{L}_{\xi}g,\phi,\delta \phi, \mathcal{L}_{\xi}\phi ) &\equiv \delta[ \mathbf{\Theta}^{(M)}(g,\phi,\mathcal{L}_{\xi} \phi)] - \mathcal{L}_{\xi}[ \mathbf{\Theta}^{(M)}(g,\phi, \delta \phi)].
\end{split}
\end{equation}
By using the Cartan's formula Eq.~\eqref{eq:Cartan}, we obtain
\begin{align}
\delta \mathbf{J}_{\xi, \text{Wald}} &=  \mathbf{\Omega}^{(G)}(\delta g, \mathcal{L}_{\xi}g)  + d(\xi \cdot \mathbf{\Theta}^{(G)}(\delta g)) +  \mathbf{\Omega}^{(M)}(\delta g,\mathcal{L}_{\xi}g,\delta \phi, \mathcal{L}_{\xi}\phi)  + d(\xi \cdot \mathbf{\Theta}^{(M)}(\delta \phi))
\quad \text{$\{g,\phi\}$ on shell}.
\end{align}

Finally we shall focus on variations $\delta g, \delta \phi$ which satisfy the linearized equations of motion. This allows us to replace $\mathbf{J}_{\xi, \text{Wald}}$ and its variation $\delta \mathbf{J}_{\xi, \text{Wald}}$ with the Noether charge $d\mathbf{Q}_{\xi, \text{Wald}}$ and its variation $\delta d\mathbf{Q}_{\xi, \text{Wald}}$. Hence, by focusing on on shell variations, we obtain 
\begin{align}
\delta d \mathbf{Q}_{\xi, \text{Wald}} =  \mathbf{\Omega}^{(G)}(\delta g, \mathcal{L}_{\xi}g) + d\big( \xi \cdot \mathbf{\Theta}^{(G)}(\delta g) \big)  + \mathbf{\Omega}^{(M)}(\delta g, \mathcal{L}_{\xi} g,\delta \phi, \mathcal{L}_{\xi}\phi) + d\big( \xi \cdot \mathbf{\Theta}^{(M)}(\delta \phi) \big) \quad \text{$\{\delta g, \delta \phi\}$ on shell}. 
\end{align}

The above expression can be further simplified when $\xi$ is a Killing vector:
\begin{align}
\mathcal{L}_{\xi} g_{\mu\nu} = 0.
\end{align}
Let us further assume a similar relation for the matter field:
\begin{align}
\mathcal{L}_{\xi} \phi = 0.
\end{align}
Then at the linear order in $\delta g$ and $\delta \phi$, we have
\begin{align}
\mathbf{\Omega}^{(G)}(\delta g, \mathcal{L}_{\xi}g)=0, \qquad \mathbf{\Omega}^{(M)}(\delta g, \mathcal{L}_{\xi}g,\delta \phi, \mathcal{L}_{\xi}\phi)=0.
\end{align}
Hence we arrive at the well-celebrated result of Wald:
\begin{align}
\delta d \mathbf{Q}_{\xi, \text{Wald}} =   d\big( \xi \cdot \mathbf{\Theta}^{(G)}(\delta g) \big) + d\big( \xi \cdot \mathbf{\Theta}^{(M)}(\delta \phi) \big),
\quad \text{$\xi$ : Killing vector, $\mathcal{L}_{\xi} \phi = 0$, $\{\delta g, \delta \phi\}$ on shell}. 
\end{align}

\subsection{Example: general relativity}

We illustrate the covariant phase space formalism and the Noether charge by looking at general relativity with scalar fields. The Lagrangian is given by
\begin{align}
\mathbf{L}^{(G)} = \frac{1}{16\pi} \pmb{\epsilon} (R-2\Lambda),
\qquad \mathbf{L}^{(M)} = - \frac{1}{2} \pmb{\epsilon}(\nabla^{\mu} \phi)(\nabla_{\mu} \phi).
\end{align}
The equations of motion are
\begin{equation}
{\mathbf{E}^{(G)}}^{\mu\nu} = - \frac{1}{16\pi}\pmb{\epsilon} G^{\mu\nu}, \qquad {\mathbf{E}^{(\phi)}} = \pmb{\epsilon} \nabla_{\mu}\nabla^{\mu} \phi
\end{equation}
where $G^{\mu\nu} \equiv R^{\mu\nu} - \frac{1}{2}R g^{\mu\nu} + \Lambda g^{\mu\nu}$.
The symplectic potentials are given by 
\begin{align}
\mathbf{\Theta}^{(G)}_{abc} = \pmb{\epsilon}_{dabc} \frac{1}{16\pi} g^{de}g^{fh} (\nabla_{f} \delta g_{eh} - \nabla_e \delta g_{fh} ), \qquad
\mathbf{\Theta}_{abc}^{(M)} = - \pmb{\epsilon}_{dabc}(\nabla^{d}\phi)\delta \phi \label{eq:GR-Theta}
\end{align}
and the energy momentum tensor is
\begin{align}
T_{\mu\nu}=\nabla_{\mu}\phi \nabla_{\nu}\phi -\frac{1}{2}g_{\mu\nu}\nabla_{\alpha}\phi \nabla^{\alpha}
\phi. 
\end{align}
For variations by diffeomorphism, we have
\begin{align}
{\mathbf{\Theta}_{\xi}}^{(G)}_{abc} = \pmb{\epsilon}_{dabc} \frac{1}{16 \pi} ( \nabla_e \nabla^d \xi^e + \nabla_e \nabla^e \xi^d - 2 \nabla^d \nabla_e \xi^e ).
\end{align}

To compute the Noether current, it is useful to make use of the following relation:
\begin{align}
\nabla_e \nabla^d \xi^e - \nabla^d \nabla_e \xi^e = - R^{\beta \; \; d \; \; }_{\; \; \alpha \; \; \beta} \xi^{\alpha} = R^{\;\; d}_{\alpha \; \;} \xi^{\alpha}.
\end{align}
We can then rewrite $\mathbf{\Theta}^{(G)}_{\xi}$ as
\begin{align}
{\mathbf{\Theta}_{\xi}}^{(G)}_{abc} = \pmb{\epsilon}_{dabc} \frac{1}{16 \pi} ( \nabla_e \nabla^e \xi^d - \nabla_e \nabla^d \xi^e + 2 R^{\; \; d}_{\alpha \; \; } \xi^{\alpha})
\end{align}
and obtain
\begin{align}
{\mathbf{J}_{\xi, \text{Wald}}}_{abc} =\pmb{\epsilon}_{dabc} \frac{1}{16 \pi} ( \nabla_e \nabla^e \xi^d - \nabla_e \nabla^d \xi^e + 2 R^{\; \; d}_{\alpha \; \; } \xi^{\alpha})- \frac{1}{16\pi} \pmb{\epsilon}_{dabc} \xi^d (R - 2\Lambda) - \pmb{\epsilon}_{dabc} T^{de} \xi_{e}.
\end{align}
By using the Einstein equation:
\begin{align}
T_{\mu\nu} = \frac{1}{8\pi}\Big(R_{\mu\nu} - \frac{1}{2}g_{\mu\nu} (R - 2\Lambda) \Big)
\end{align}
we arrive at 
\begin{align}
{\mathbf{J}_{\xi, \text{Wald}}}_{abc} =\pmb{\epsilon}_{dabc} \frac{1}{16 \pi} ( \nabla_e \nabla^e \xi^d - \nabla_e \nabla^d \xi^e). 
\end{align}

Hence the expression of $\mathbf{J}_{\xi, \text{Wald}}$ is written entirely with the metric and does not contain $\Lambda$, $\phi$ or $T_{\mu\nu}$ explicitly. Finally we obtain the Noether charge
\begin{align}
{\mathbf{Q}_{\xi, \text{Wald}}}_{ab} = - \frac{1}{16\pi} \pmb{\epsilon}_{abcd}\nabla^c \xi^d.
\end{align}


\section{Two refinements to Wald's entropy formula}\label{sec:refinement}

The original Noether charge method is not particularly suitable for studying the gravitational shock wave geometries. In this section, we will make two refinements to the Wald's entropy formula.

First, we will derive an alternative expression of the Noether current $\mathbf{J}_{\xi, \text{field}}$, which differs from the Wald's construction $\mathbf{J}_{\xi, \text{Wald}}$, by evaluating the Lie derivative of the matter Lagrangian $\mathcal{L}_{\xi} \mathbf{L}^{(M)}$ explicitly and relating it to the energy momentum tensor $T_{\mu\nu}$. Our construction of the Noether charge $\mathbf{Q}_{\xi, \text{field}}$ is particularly useful for studying the gravitational backreaction as it relates the energy momentum tensor variation directly to the metric variation. 

Second, we will extend the linear order analysis by Wald to the higher order matter field variations $\delta \phi$ while keeping the metric variation $\delta g$ at the linear order. This is due to the difficulty that contributions from the linear order variation $\delta\phi$ vanish for the gravitational shock wave geometries. Our construction of the Noether current $\mathbf{Q}_{\xi, \text{field}}$ enables us to evaluate its variation $\delta \mathbf{Q}_{\xi, \text{field}}$ under the matter field variation $\delta \phi$ at \emph{any order} in a systematic manner. 

\subsection{Matter current}

Our derivation deviates from Wald's where we evaluate the Lie derivative $\mathcal{L}_{\xi} \mathbf{L}^{(M)}$ of the matter Lagrangian in Eq.~\eqref{eq:Lie-matter}. Diffeomorphism variation of the matter Lagrangian can be explicitly evaluated as follows:
\begin{equation}
\begin{split}
\hat{\delta}_{\xi} \mathbf{L}^{(M)}&= \mathcal{L}_{\xi} \mathbf{L}^{(M)} =\mathcal{L}_{\xi} g_{\mu\nu}
\frac{\delta \mathbf{L}^{(M)}}{\delta g_{\mu\nu}}+
\mathcal{L}_{\xi}\phi \frac{\delta \mathbf{L}^{(M)}}{\delta \phi}+
d\mathbf{\Theta}_{\xi}^{(M)}\\
&=2\nabla_{(\mu}\xi_{\nu)}\frac{\delta \mathbf{L}^{(M)}}{\delta g_{\mu\nu}}+
\mathcal{L}_{\xi}\phi \frac{\delta \mathbf{L}^{(M)}}{\delta \phi}+
d\mathbf{\Theta}_{\xi}^{(M)}\\
&=\mathbf{d^dx}\sqrt{-g}
T^{\mu\nu}\nabla_{(\mu}\xi_{\nu)}+\mathcal{L}_{\xi}\phi \frac{\delta \mathbf{L}^{(M)}}{\delta \phi}+
d\mathbf{\Theta}_{\xi}^{(M)}\\
&= \mathbf{d^dx}\sqrt{-g} \nabla_{\mu}(T^{\mu\nu}\xi_{\nu})-\mathbf{d^dx}\sqrt{-g}(\nabla_{\mu}T^{\mu\nu})\xi_{\nu}+\mathcal{L}_{\xi}\phi \ \mathbf{E}^{(\phi)}+
d\mathbf{\Theta}_{\xi}^{(M)}
\\&=- \pmb{\epsilon} (\nabla_{\mu} T^{\mu\nu})\xi_{\nu} + \mathbf{d^d x} \partial_{\mu} (\sqrt{-g} T^{\mu\nu}\xi_{\nu}) +d \mathbf{\Theta}_{\xi}^{(M)} 
\end{split}
\label{eq:M-variation}
\end{equation}
where we made use of the metric compatibility 
\begin{align}
\nabla_{\xi} g_{\mu\nu} =0, \qquad \mathcal{L}_{\xi} g_{\mu\nu}  = \nabla_{\mu}\xi_{\nu} + \nabla_{\nu}\xi_{\mu}
\end{align}
the evaluation of the integral by parts, and on shell condition of matter fields. Let us define the matter current $(d-1)$-form by 
\begin{align}
\boxed{ \ 
{\mathbf{J}_{\xi, \text{field}}^{(M)}} \equiv  {J^{(M)}_{\xi, \text{field}}}^{\mu}\sqrt{-g} (\mathbf{d^{d-1} x})_{\mu} , \qquad {J^{(M)}_{\xi, \text{field}}}^{\mu} \equiv - T^{\mu\nu}\xi_{\nu}.
\ }
\end{align}
Here $(\mathbf{d^{d-1} x})_{\mu}$ is a $(d-1)$-form that does \emph{not} contain $\mathbf{dx^\mu}$
~\footnote{
Explicitly, we have 
\begin{align}
 (\mathbf{d^{d-p}}x)_{\mu_1 \dots \mu_p}=\frac{1}{p!(n-p)!}\epsilon_{\mu_1\dots\mu_p\nu_{p+1}\dots\nu_d}\ \mathbf{dx}^{\nu_{p+1}}\wedge \dots\wedge \mathbf{dx}^{\nu_d}
\end{align}
where $\epsilon$ is the Levi-Civita symbol. Here ${\mathbf{J}_{\xi}^{(M)}}$ should be understood as the Hodge dual of the $1$-form current.
}. 
We then obtain 
\begin{equation}
\begin{split}
\mathcal{L}_{\xi} \mathbf{L}^{(M)} &= - \pmb{\epsilon} (\nabla_{\mu} T^{\mu\nu})\xi_{\nu} - d\mathbf{J}_{\xi, \text{field}}^{(M)} +d \mathbf{\Theta}_{\xi}^{(M)} \qquad \text{$\phi$ on shell}.\label{eq:45}
\end{split}
\end{equation}

Recall that $\mathcal{L}_{\xi} \mathbf{L}^{(M)}$ can be expressed as a total derivative, according to Eq.~\eqref{eq:Lie-matter}. Thus, from Eq.~\eqref{eq:45}, we can deduce the conservation of the energy momentum tensor:
\begin{align}
\nabla_{\mu} T^{\mu\nu} = 0 \qquad \text{$\phi$ on shell}.
\end{align}
Then, the Lie derivative of the matter Lagrangian $\mathcal{L}_{\xi} \mathbf{L}^{(M)}$ can be expressed as
\begin{align}
\mathcal{L}_{\xi} \mathbf{L}^{(M)} = -  d\mathbf{J}_{\xi, \text{field}}^{(M)}+ d \mathbf{\Theta}_{\xi}^{(M)} \qquad \text{$\phi$ on shell}.
\end{align}
Also, the Lie derivative of the gravity Lagrangian is given by
\begin{align}
\mathcal{L}_{\xi} \mathbf{L}^{(G)} =  d\mathbf{J}_{\xi, \text{field}}^{(M)}+ d \mathbf{\Theta}_{\xi}^{(G)}\qquad \text{$\{ g, \phi\}$ on shell}. \label{eq:G-variation2}
\end{align}
Here it is worth emphasizing that $\mathcal{L}_{\xi} \mathbf{L}^{(G)}\not=d \mathbf{\Theta}_{\xi}^{(G)}$ and $\mathcal{L}_{\xi} \mathbf{L}^{(M)}\not=d \mathbf{\Theta}_{\xi}^{(M)}$ due to the matter current. 

We define the Noether current $(d-1)$-form by 
\begin{align}
\boxed{ \ 
\mathbf{J}_{\xi, \text{field}} \equiv \mathbf{\Theta}^{(G)}_{\xi} - \xi \cdot \mathbf{L}^{(G)}  + \mathbf{J}_{\xi,  \text{field}}^{(M)}. \ }
 \label{eq:Noether-current-QFT}
\end{align}
One can verify that $\mathbf{J}_{\xi, \text{field}} $ is closed when $\{g,\phi\}$ are on shell:
\begin{equation}
\begin{split}
d \mathbf{J}_{\xi,\text{field}} &= d \mathbf{\Theta}^{(G)}_{\xi} - d(\xi \cdot \mathbf{L}^{(G)}) + d\mathbf{J}_{\xi,  \text{field}}^{(M)} \\
&= d \mathbf{\Theta}^{(G)}_{\xi} - \mathcal{L}_{\xi} \mathbf{L}^{(G)}  + d\mathbf{J}_{\xi,  \text{field}}^{(M)} \\
&=0.
\end{split}
\end{equation}
Then one can construct the Noether charge $\mathbf{Q}_{\xi, \text{field}}$:
\begin{align}
\exists \mathbf{Q}_{\xi, \text{field}} \ \ \text{such that} \ \ \mathbf{J}_{\xi, \text{field}} = d \mathbf{Q}_{\xi, \text{field}} \qquad \text{$\{g, \phi\} $ on shell}.
\end{align}
Note that our definition $\mathbf{J}_{\xi, \text{field}}$ differs from Wald's $\mathbf{J}_{\xi, \text{Wald}}$ since, instead of $\mathbf{\Theta}^{(M)}_{\xi} - \xi \cdot \mathbf{L}^{(M)} $, we have used $\mathbf{J}_{\xi, \text{field}}^{(M)}$. We will compare two constructions in section~\ref{sec:comparison}

\subsection{Higher-order matter variation}

Until this point, our derivation of the Noether current Eq.~\eqref{eq:Noether-current-QFT} is an exact calculation which does not rely on perturbative analysis~\footnote{
Recall that it suffices to use the linearized variation formula Eq.~\eqref{eq:variation} to evaluate the Lie derivative. 
}. The remaining task is to evaluate the variation of the Noether current Eq.~\eqref{eq:Noether-current-QFT} under $\delta g_{\mu\nu}, \delta \phi$. 

We will focus on the linear order metric variation $\delta g_{\mu\nu}$ while we do not impose such restriction on the matter field variation $\delta \phi$. We then obtain
\begin{equation}
\begin{split}
\delta \mathbf{J}_{\xi, \text{field}} &= \delta[\mathbf{\Theta}^{(G)}(\mathcal{L}_{\xi}g)] - \xi\cdot \delta \mathbf{L}^{(G)}  + \delta\mathbf{J}_{\xi, \text{field}}^{(M)}\\
&= \delta[\mathbf{\Theta}^{(G)}(\mathcal{L}_{\xi}g)] - \xi \cdot d \mathbf{\Theta}^{(G)}(\delta g)  + \delta \mathbf{J}_{\xi, \text{field}}^{(M)} -\xi\cdot \mathbf{E}^{(G)\mu\nu}\delta g_{\mu\nu} \\ 
&=  \delta[\mathbf{\Theta}^{(G)}(\mathcal{L}_{\xi}g)] - \mathcal{L}_{\xi}[\mathbf{\Theta}^{(G)}(\delta g) ] + d(\xi \cdot \mathbf{\Theta}^{(G)}(\delta g) ) + \delta\mathbf{J}_{\xi, \text{field}}^{(M)} -\xi\cdot \mathbf{E}^{(G)\mu\nu}\delta g_{\mu\nu}\\
&=  \delta[\mathbf{\Theta}^{(G)}(\mathcal{L}_{\xi}g)] - \mathcal{L}_{\xi}[\mathbf{\Theta}^{(G)}(\delta g) ] + d(\xi \cdot \mathbf{\Theta}^{(G)}(\delta g) ) + \delta\mathbf{J}_{\xi, \text{field}}^{(M)} +\frac{1}{2}(\xi\cdot \pmb{\epsilon})T^{\mu\nu}\delta g_{\mu\nu}
 \label{eq:derivation}
\end{split}
\end{equation}
where we used $\mathbf{E}^{(G)\mu\nu}= -\frac{1}{2} \pmb{\epsilon}T^{\mu\nu}$. 
By using the gravity part of the symplectic current $\mathbf{\Omega}^{(G)}(\delta g, \mathcal{L}_{\xi}g ) \equiv \delta[ \mathbf{\Theta}^{(G)}(\mathcal{L}_{\xi} g)] - \mathcal{L}_{\xi}[ \mathbf{\Theta}^{(G)}(\delta g)]$, we obtain
~\footnote{In fact, we do not need the on shell condition for the matter fields $\phi$ for this equation to hold. We would need it to make sure $d\mathbf{J}_{\xi, \text{field}}^{(M)}=0$ so that the matter charge $\mathbf{Q}_{\xi, \text{field}}^{(M)}$ can be constructed.}
\begin{align}
\delta \mathbf{J}_{\xi, \text{field}} &=  \mathbf{\Omega}^{(G)}(\delta g, \mathcal{L}_{\xi}g)  + d(\xi \cdot \mathbf{\Theta}^{(G)}(\delta g)) + \delta\mathbf{J}_{\xi, \text{field}}^{(M)}+\frac{1}{2}(\xi\cdot \pmb{\epsilon})T^{\mu\nu}\delta g_{\mu\nu}
\qquad \text{$\{g,\phi\}$ on shell}.  \label{eq:J-variation-QFT}
\end{align}
It is worth emphasizing again that Eq.~\eqref{eq:J-variation-QFT} is valid up to the linear order in $\delta g_{\mu\nu}$ and up to \emph{any order} in $\delta \phi$. Note that the matter field variation $\delta\phi$ enters through the matter current variation $ \delta\mathbf{J}_{\xi, \text{field}}^{(M)}$. Finally we shall focus on variations $\delta g, \delta \phi$ which satisfy the equations of motion up to the linear order in $\delta g$ and up to any desired order in $\delta \phi$. 

Finally, we arrive at 
\begin{align}
\boxed{\
\delta d \mathbf{Q}_{\xi, \text{field}} =  \mathbf{\Omega}^{(G)}(\delta g, \mathcal{L}_{\xi}g) + d\big( \xi \cdot \mathbf{\Theta}^{(G)}(\delta g) \big) + \delta \mathbf{J}^{(M)}_{\xi, \text{field}}+\frac{1}{2}(\xi\cdot\pmb{ \epsilon}) T^{\mu\nu} \delta g_{\mu\nu} \qquad \text{$\{\delta g, \delta \phi\}$ on shell}. \label{eq:Noether}\ }
\end{align}
We will use this formula to derive the microscopic thermodynamic relation.

The above expression can be further simplified when $\xi$ is a Killing vector of the spacetime:
\begin{align}
\mathcal{L}_{\xi} g_{\mu\nu} = 0.
\end{align}
Then we have $\mathbf{\Omega}^{(G)}(\delta g, \mathcal{L}_{\xi}g)=0$ in the linear order in $\delta g$.

When $\xi$ is a Killing vector, the matter current is conserved:
\begin{align}
\nabla_\mu {J^{(M)}_{\xi, \text{field}}}^{\mu} = -\nabla_\mu (T^{\mu\nu}\xi_\nu) = - (\nabla_\mu T^{\mu\nu}) \xi_\nu - T^{\mu\nu} \nabla_\mu\xi_\nu = - \frac{1}{2}T^{\mu\nu} (\nabla_\mu\xi_\nu + \nabla_\nu\xi_\mu)=0
\end{align}
or equivalently
\begin{align}
d {\mathbf{J}_{\xi, \text{field}}^{(M)}} = 0. 
\end{align}
Thus, we can construct the matter charge $(d-2)$-form $\mathbf{Q}_{\xi, \text{field}}^{(M)}$ such that
\begin{align}
{\mathbf{J}_{\xi, \text{field}}^{(M)}}  = d  {\mathbf{Q}_{\xi, \text{field}}^{(M)}}. 
\end{align}
Hence we arrive at the following expression:
\begin{align}
\delta d \mathbf{Q}_{\xi, \text{field}} =   d\big( \xi \cdot \mathbf{\Theta}^{(G)}(\delta g) \big) + \delta d \mathbf{Q}^{(M)}_{\xi, \text{field}} +\frac{1}{2}(\xi\cdot \pmb{\epsilon})T^{\mu\nu}\delta g_{\mu\nu},
\quad \text{$\xi$ : Killing vector, $\{\delta g, \delta \phi\}$ on shell} \label{eq:Killing}
\end{align}
which is valid up to linear order in $\delta g$ and any order in $\delta \phi$. 

\subsection{Comparison}\label{sec:comparison}

Let us explicitly compare two possible constructions of the matter current
\begin{equation}
\begin{split}
\mathbf{J}^{(M)}_{\xi, \text{Wald}} &\equiv \mathbf{\Theta}_{\xi}^{(M)}(\phi,\mathcal{L}_{\xi}\phi)-\xi\cdot \mathbf{L}^{(M)} \\
{\mathbf{J}_{\xi, \text{field}}^{(M)}} &\equiv  {J^{(M)}_{\xi, \text{field}}}^{\mu}\sqrt{-g} (\mathbf{d^{d-1} x})_{\mu} , \qquad \text{where} \quad {J^{(M)}_{\xi, \text{field}}}^{\mu} \equiv - T^{\mu\nu}\xi_{\nu}.
\end{split}
\end{equation}
To begin, we compare their total derivatives:
\begin{equation}
\begin{split}
d\mathbf{J}^{(M)}_{\xi, \text{Wald}} &=d\mathbf{\Theta}_{\xi}^{(M)}-d(\xi\cdot \mathbf{L}^{(M)})\\
&=-\mathbf{E}^{(\phi)}\mathcal{L}_{\xi}\phi-\frac{1}{2}\pmb{\epsilon}T^{\mu\nu} \mathcal{L}_{\xi}g_{\mu\nu}
\end{split}
\end{equation}
and
\begin{equation}
\begin{split}
d\mathbf{J}^{(M)}_{\xi, \text{field}}&=-\mathbf{d^d x}\sqrt{-g}\nabla_{\mu}(T^{\mu\nu}\xi_{\nu})\\
&=-\pmb{\epsilon} \left( (\nabla_{\mu}T^{\mu\nu})\xi_{\nu}+T^{\mu\nu}\nabla_{(\mu}\xi_{\nu)}\right)\\
&=-\pmb{\epsilon} \left( (\nabla_{\mu}T^{\mu\nu})\xi_{\nu}+\frac{1}{2}T^{\mu\nu}\mathcal{L}_{\xi}g_{\mu\nu}\right).
\end{split}
\end{equation}
Their difference indeed vanishes for matter fields on shell:
\begin{align}
d\mathbf{J}^{(M)}_{\xi,\text{field}}-d\mathbf{J}^{(M)}_{\xi,\text{Wald}}=\mathbf{E}^{(\phi)}\mathcal{L}_{\xi}\phi-\pmb{\epsilon}(\nabla_{\mu}T^{\mu\nu})\xi_{\nu} = 0 \qquad \mbox{$\phi$ on shell}.
\end{align}
Hence both $\mathbf{J}^{(M)}_{\xi,\text{Wald}}$ and $\mathbf{J}^{(M)}_{\xi,\text{field}}$ lead to valid constructions of the Noether charge $\mathbf{Q}_{\xi,\text{Wald}}$ and $\mathbf{Q}_{\xi,\text{field}}$. 

Next, let us look at the scalar field Lagrangian 
\begin{align}
\mathbf{L}_{\text{scalar}}=-\frac{1}{2}\pmb{\epsilon}g^{\mu\nu}\nabla_{\mu}\phi \nabla_{\nu}\phi.
\end{align}
We have
\begin{equation}
\begin{split}
\mathbf{J}^{(M)}_{\xi,\text{field}}&=-T^{\mu\nu}\xi_{\nu}\sqrt{-g}(\mathbf{d^{d-1}x})_{\mu}\\
&=-(\nabla^{\mu}\phi\nabla^{\nu}\phi\  \xi_{\nu}-\frac{1}{2}\xi^{\mu}\nabla_{\alpha}\phi \nabla^{\alpha} \phi)\sqrt{-g}(\mathbf{d^{d-1}x})_{\mu}\\
\mathbf{J}^{(M)}_{\xi, \text{Wald}}
&=-\big( (\nabla^\mu \phi) \mathcal{L}_{\xi}\phi-\frac{1}{2}\xi^\mu \nabla_{\alpha}\phi \nabla^{\alpha}\phi
\big) \sqrt{-g}(\mathbf{d^{d-1}x})_{\mu} \\
&=- \big( \xi_{\nu}\nabla^\nu \phi\nabla^{\mu}\phi-\frac{1}{2}\xi^\mu \nabla_{\alpha}\phi \nabla^{\alpha}\phi
\big) \sqrt{-g}(\mathbf{d^{d-1}x})_{\mu} . 
\end{split}
\end{equation}  
Hence, we have
\begin{align}
\mathbf{J}^{(M)}_{\xi,\text{field}} = \mathbf{J}^{(M)}_{\xi,\text{Wald}}.
\end{align} 

For the electromagnetic Lagrangian
\begin{align}
\mathbf{L}_{\text{EM}}=-\frac{1}{4}\pmb{\epsilon}F_{\mu\nu}F^{\mu\nu}
\end{align}
we have
\begin{align}
T_{\mu\nu}=F_{\mu\alpha}F_{\nu}^{\;\;\alpha}-\frac{1}{4}g_{\mu\nu}F_{\alpha\beta}F^{\alpha\beta}
\\
 \mathbf{\Theta}^{(M)}_{abc}(A,\delta A)=-\pmb{\epsilon}_{\mu abc}F^{\mu \alpha}\delta A_{\alpha}
 \\
 \mathbf{E}^{(A),\alpha}=\frac{\delta \mathbf{L}_{\text{EM}}}{\delta A_{\alpha}}=\pmb{\epsilon}\nabla_{\mu}F^{\mu \alpha}
\end{align}
and
\begin{equation}
\begin{split}
\mathbf{J}^{(M)}_{\xi,\text{field}}&=-T^{\mu\nu}\xi_{\nu}\sqrt{-g}(\mathbf{d^{d-1}x})_{\mu}\\
&=-(F^{\mu}_{\;\;\alpha}F^{\nu\alpha}-\frac{1}{4}g^{\mu\nu}F_{\alpha\beta}F^{\alpha \beta})\xi_{\nu}\sqrt{-g}(\mathbf{d^{d-1}x})_{\mu}\\
\mathbf{J}^{(M)}_{\xi,\text{Wald}}
&=-(F^{\mu \alpha}\mathcal{L}_{\xi}A_{\alpha}-\frac{1}{4}\xi^{\mu}F_{\alpha\beta}F^{\alpha\beta})\sqrt{-g}(\mathbf{d^{d-1}x})_{\mu}\\
&=-(F^{\mu \alpha}\xi^{\nu}F_{\nu\alpha}+F^{\mu \alpha}\nabla_{\alpha}(\xi^{\nu}A_{\nu})-\frac{1}{4}\xi^{\mu}F_{\alpha\beta}F^{\alpha\beta})\sqrt{-g}(\mathbf{d^{d-1}x})_{\mu}
\end{split}
\end{equation}  
where we used the Cartan's magic formula for the Lie derivative. Hence we obtain 
\begin{align}
\mathbf{J}^{(M)}_{\xi,\text{field}}-\mathbf{J}^{(M)}_{\xi,\text{Wald}}= F^{\mu\alpha}\nabla_{\alpha}(\xi^{\nu}A_{\nu})\sqrt{-g}(\mathbf{d^{d-1}x})_{\mu}
\end{align} 
which, by imposing on shell conditions $\nabla_{\alpha}F^{\mu \alpha}=0$, becomes
\begin{align}
\mathbf{J}^{(M)}_{\xi,\text{field}}-\mathbf{J}^{(M)}_{\xi,\text{Wald}}= \nabla_{\alpha}(F^{\mu\alpha}\xi^{\nu}A_{\nu})\sqrt{-g}(\mathbf{d^{d-1}x})_{\mu} \qquad (A_{\nu}\ \text{on shell}). 
\end{align}

Note that two definitions of the matter current for the electromagnetic field differ by a total derivative. Thus we can write 
\begin{align}
\mathbf{J}^{(M)}_{\xi,\text{field}}-\mathbf{J}^{(M)}_{\xi,\text{Wald}}= d \mathbf{Q}^{(\Delta)}_{\xi}
\end{align} 
which leads to the following difference of two constructions:
\begin{align}
\mathbf{Q}_{\xi,\text{field}}-\mathbf{Q}_{\xi,\text{Wald}}=  \mathbf{Q}^{(\Delta)}_{\xi}.
\end{align}
Here it is worth emphasizing that the charge difference $\mathbf{Q}^{(\Delta)}_{\xi}$ is linear in the vector field $\xi$, and thus vanishes as $\xi\rightarrow 0$.

In fact, by following Iyer and Wald (Lemma 3.1 in~\cite{Iyer:1994ys}), one can show that, for an arbitrary matter Lagrangian, 
\begin{enumerate}[(a)]
\item $\mathbf{J}^{(M)}_{\xi,\text{field}}$ and $\mathbf{J}^{(M)}_{\xi,\text{Wald}}$ differ only by a total derivative. 
\item The charge difference $\mathbf{Q}^{(\Delta)}_{\xi}$ vanishes as $\xi \rightarrow 0$. 
\end{enumerate}
For our application to the gravitational shock wave geometries, terms which vanish as $\xi\rightarrow 0$ do not contribute to the surface integral of the Noether charge. As such, our construction of the Noether charge $\mathbf{Q}_{\xi,\text{field}}$ bears the same geometrical meaning as the Wald's construction $\mathbf{Q}_{\xi,\text{Wald}}$.


\section{Gravitational shock wave}
\label{sec:shockwave}

Having reviewed and refined the Noether charge formula, let us shift the gear a bit. In this section, we present a brief review of gravitational shock wave geometries by following the works by Dray, 't Hooft and Sfetsos \cite{Dray:1984ha, Sfetsos:1994xa} who derived the gravitational shock wave equations of motion from the Einstein equation. 

Consider the following family of $d$-dimensional static spacetime, expressed in the Kruskal-type coordinate and its corresponding energy momentum tensor:
\begin{equation}
\begin{split}
ds^2 &= 2 F(u,v) dudv + G(u,v) h_{ij} dx^i dx^j \\ 
 T &= 2T_{uv}dudv + T_{uu}dudu + T_{vv}dvdv  + T_{ij}dx^{i}dx^{j}. \label{eq:original}
\end{split}
\end{equation}
Following Sfetsos~\cite{Sfetsos:1994xa}, we assume that the metric and matter fields satisfy the following conditions:
\begin{align}
&G_{,v} = F_{,v} = T_{vv} =0 \qquad (u=0) \label{eq:u-assumption} \\
&G_{,u} = F_{,u} = T_{uu} =0 \qquad (v=0) \label{eq:v-assumption}.
\end{align}
The above conditions follow from the assumption that the timelike vector $\partial_t$ is a Killing vector of the spacetime as we will see in the next subsection. Non-vanishing Christoffel symbols are
\begin{equation}
\begin{split}
&\Gamma^u_{uu} = \frac{F_{,u}}{F} \qquad \Gamma^u_{ij}= -\frac{G_{,v}}{2F}\ h_{ij} \qquad
\Gamma^v_{vv} = \frac{F_{,v}}{F} \qquad \Gamma^v_{ij}= -\frac{G_{,u}}{2F}\ h_{ij}  \\
&\Gamma^i_{uj} = \frac{G_{,u}}{2G}\delta^{i}_{j} \qquad \Gamma^i_{vj} = \frac{G_{,v}}{2G}\delta^{i}_{j} \qquad
\Gamma^{i}_{jk} = \frac{1}{2}h^{il}( h_{lk,j} + h_{lj,k} - h_{jk,l}). \label{eq:symbol}
\end{split}
\end{equation}

\begin{figure}
\centering
\includegraphics[width=0.35\textwidth]{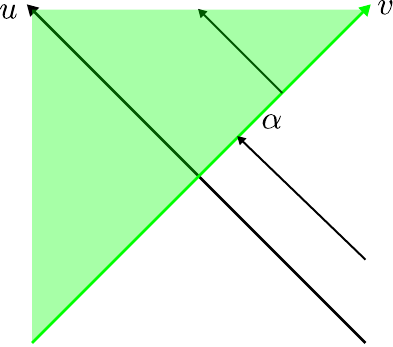} 
\caption{Gravitational shock wave at $u=0$, moving in the $v$-direction. The shaded region experiences metric variations.
}
\label{fig-shock}
\end{figure}
Let us shift this spacetime by 
\begin{align}
v \quad \rightarrow \quad \tilde{v}=v + \theta(u)\alpha(x)
\end{align}
where $x$ represents $x^j$ collectively and $\theta(u)$ is a step function. The resulting metric and energy momentum tensor are 
\begin{align}
d\tilde{s}^2 = 2 F(u,v+\theta \alpha) du(dv + \theta \alpha_{,i} dx^i ) + G(u,v+\theta \alpha) h_{ij} dx^i dx^j \label{eq:shockwave}
\end{align}
and
\begin{equation}
\begin{split}
\tilde{T}&=2T_{uv}(u, v+\theta \alpha, x)du(dv+\theta \alpha_{,i} dx^i)+T_{uu}(u,v+\theta \alpha,x)du^2\\
&+T_{vv}(u,v+\theta \alpha,x)(dv+\theta \alpha_{,i} dx^i)^2+T_{ij}(u,v+\theta \alpha,x)dx^idx^j. \label{eq:shockwave2}
\end{split}
\end{equation}

Note that the shifted metric and the shifted energy momentum tensor still satisfy the Einstein equation except at $u=0$ where an additional source of the energy momentum tensor is needed. Consider the extra contribution to the energy momentum tensor from a massless particle, located at $u=0$ and moving with the speed of light in the $v$-direction:
\begin{align}
T^{(P)} = T_{uu}^{(P)} du^2 =   4p^{(v)} F^2 \delta(x) \delta(u) du^2 \label{eq:point-particle}
\end{align}
where $p^{(v)}$ is the momentum of the particle; $T^{vv}= 4p^{(v)}\delta(u)\delta(x)$
~\footnote{
Our choice of $T^{vv}= 4p^{(v)}\delta(u)\delta(x)$ differs from Sfetsos' by a factor of $(-1)$ due to the sign difference in the Einstein equation.
}. The Einstein equation is satisfied if Eq.~\eqref{eq:u-assumption} holds \emph{and} the following equation holds:
\begin{align}
\Delta_{h_{ij}} \alpha(x) - \frac{d-2}{2} \frac{G_{,uv}}{F}  \alpha(x) = 32 \pi p^{(v)} G F \delta(x). 
 \label{eq:shock}
\end{align}
Here the Laplacian is defined as $\Delta_{h_{ij}} \equiv \frac{1}{\sqrt{h}} \partial_i \sqrt{h} h^{ij}\partial_j$. We will derive this equation by using the Noether charge method later~\footnote{We note that Eq.~\eqref{eq:v-assumption} is not necessary to derive Eq.~\eqref{eq:shock}. We will however assume both Eq.~\eqref{eq:u-assumption} and Eq.~\eqref{eq:v-assumption} in this paper since we are interested in static spacetime.}. 


\section{Horizon translation vector}\label{sec:Killing}

In this section, we investigate the symmetry properties of the spacetime. The gravitational shock wave solutions require certain consistency conditions on the fields $g_{\mu\nu}, T_{\mu\nu}$ as we reviewed in the previous section. These conditions actually follow from the existence of the timelike Killing vector $\xi^{(t)}=\partial_t$ as we shall show in this section. We will also study the effect of a vector field $\hat{\beta}$ which introduce the horizon translation. In later sections, we will use the horizon translation vector $\hat{\beta}$ to construct the Noether charge which is sensitive to the gravitational shock wave.

\subsection{Timelike vector}

In this subsection, we verify that Eq.~\eqref{eq:u-assumption} and Eq.~\eqref{eq:v-assumption} follow from the fact that the timelike  $\xi^{(t)}=\partial_t$ vector is a Killing vector.

A vector field $\xi$ is a Killing vector if $\mathcal{L}_{\xi} g_{\mu\nu}=0$. Let us begin by computing $\mathcal{L}_{\xi} g_{\mu\nu}$ for the metric given in Eq.~\eqref{eq:original}. Looking at $(\mu,\nu)=(u,v)$, we have 
\begin{equation}
\begin{split}
\mathcal{L}_{\xi}g_{uv} &= \xi^a (\partial_a g_{uv}) + (\partial_u \xi^a)g_{av} + (\partial_v \xi^a)g_{ua} \\
&= \xi^u F_{,u} + \xi^v F_{,v} + (\partial_u\xi^u + \partial_v\xi^v) F.
\end{split}
\end{equation}
Looking at $(\mu,\nu)=(i,j)$, we have 
\begin{equation}
\begin{split}
\mathcal{L}_{\xi}g_{ij} &= \xi^a (\partial_a g_{ij}) + (\partial_i \xi^a)g_{aj} + (\partial_j \xi^a)g_{ia} \\
&= \xi^u G_{,u} h_{ij} + \xi^v G_{,v} h_{ij} + \xi^k (\partial_k g_{ij}) + (\partial_i \xi^k)g_{kj}+ (\partial_j \xi^k)g_{ik}.
\end{split}
\end{equation}
Looking at $(\mu,\nu)=(u,i)$, we have 
\begin{equation}
\begin{split}
\mathcal{L}_{\xi}g_{ui} &= \xi^a (\partial_a g_{ui}) + (\partial_u \xi^a)g_{ai} + (\partial_i \xi^a)g_{ua} \\
&= (\partial_u \xi^j)g_{ji}+ (\partial_i \xi^v)g_{uv}.
\end{split}
\end{equation}
Looking at $(\mu,\nu)=(v,i)$, we have 
\begin{equation}
\begin{split}
\mathcal{L}_{\xi}g_{vi} &= \xi^a (\partial_a g_{vi}) + (\partial_v \xi^a)g_{ai} + (\partial_i \xi^a)g_{va} \\
&= (\partial_v \xi^j)g_{ji}+ (\partial_i \xi^u)g_{uv}.
\end{split}
\end{equation}

\begin{figure}
\centering
(a)\includegraphics[width=0.35\textwidth]{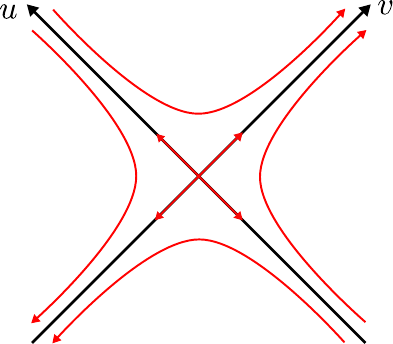} \qquad
(b)\includegraphics[width=0.35\textwidth]{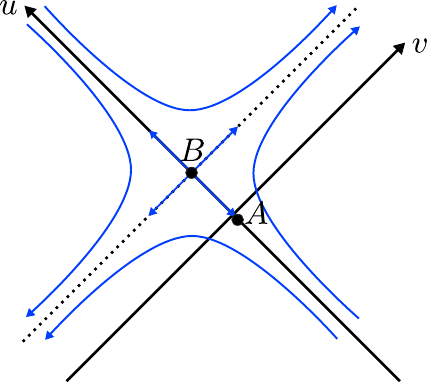} 
\caption{(a) The timelike Killing vector $\xi^{(t)}$. (b) The shifted timelike vector $\chi$.
}
\label{fig-Killing}
\end{figure}

Consider the following timelike vector field:
\begin{align} 
\xi^{(t)} \equiv \kappa (-u\partial_u + v \partial_v) 
\end{align}
where $\kappa$ represents the surface gravity. One can verify 
\begin{align}
&\mathcal{L}_{\xi^{(t)}}g_{uv},\ \mathcal{L}_{\xi^{(t)}}g_{ij} \rightarrow  0\qquad  \text{ $u \rightarrow  0$ or  $v \rightarrow  0$}
\end{align}
and $\mathcal{L}_{\xi^{(t)}}g_{ui}, \mathcal{L}_{\xi^{(t)}}g_{vi} =0$, suggesting that $\xi^{(t)}$ asymptotically becomes a Killing vector on the horizon at $u=0$ and $v=0$ due to Eq.~\eqref{eq:u-assumption}~\eqref{eq:v-assumption}. Furthermore, we see that the norm of $\xi^{(t)}$ vanishes on the horizon, and $\xi^{(t)}=0$ at $(u,v)=(0,0)$~\footnote{
The norm is given by $2\xi^{u}g_{uv}\xi^v = - 2\kappa^2 uv F \rightarrow 0$ for $u,v \rightarrow 0$ where we assumed $F(u,v)$ is not divergent at $u=0$ or $v=0$. For non-static black holes, such as the Kerr black hole, $g_{uv}$ in the Kruskal-like coordinate contains a term proportional to $(uv)^{-1}$. As such, $u=0$ or $v=0$ hypersurfaces are not horizons with respect to the timelike Killing vector $\partial_t$.
}. Hence, $u=0$ and $v=0$ hypersurfaces are Killing horizons of $\xi^{(t)}$. In static black holes such as the Schwarzschild black hole, $\xi^{(t)}$ corresponds exactly to the timelike Killing vector $\partial_t$~\footnote{As for $T_{\mu\nu}$, we were not able to verify $\mathcal{L}_{\xi}T_{\mu\nu}=0$ from the conditions $T_{vv}=0$ ($u=0$) and $T_{uu}=0$ ($v=0$) in Eqs.~\eqref{eq:u-assumption}\eqref{eq:v-assumption}.  Note, however, that we did not need to impose $\mathcal{L}_{\xi}\phi=0$ or $\mathcal{L}_{\xi} T_{\mu\nu}=0$ in deriving the Noether charge relation in section~\ref{sec:refinement}.}.

\subsection{Horizon translation}

Next, consider the following horizon translation vector field:
\begin{align}
\hat{\beta} \equiv  \beta(x)   \partial_u 
\label{eq:shifted-vector}
\end{align}
where $\beta(x)$ is an arbitrary function of $x^i$. With direct calculations, one finds 
\begin{equation}
\begin{split}
\mathcal{L}_{\hat{\beta}}g_{uv} &= \beta(x) \big( \partial_u F \big),  \qquad
\mathcal{L}_{\hat{\beta}}g_{ij} = \beta(x) h_{ij}\big( \partial_u G \big)\\
\mathcal{L}_{\hat{\beta}}g_{vi} &= \big(\partial_i \beta(x) \big) F, \qquad
\mathcal{L}_{\hat{\beta}}g_{ui} =0.
\end{split}
\end{equation}
Namely, at the limit of $v\rightarrow 0$, we have 
\begin{equation}
\begin{split}
\mathcal{L}_{\hat{\beta}}g_{uv},\mathcal{L}_{\hat{\beta}}g_{ij},  \mathcal{L}_{\hat{\beta}}g_{ui} \rightarrow 0, \qquad
\mathcal{L}_{\hat{\beta}}g_{vi} = \big(\partial_i \beta(x) \big) F \not\rightarrow 0. 
\end{split}
\end{equation}

In the next section, we will use this horizon shift vector $\hat{\beta}$ to construct the Noether charge and derive the microscopic thermodynamic relations as in Eq.~\eqref{eq:summary}. 
It is worth recalling that, in the original treatment by Wald, the vector field $\xi$ in the Noether charge $\mathbf{Q}_{\xi}$ was chosen to be a Killing vector. 
This was crucial in the derivation of Wald's Noether charge relation since the pre-symplectic form  $\mathbf{\Omega}^{(G)}(g,\delta g, \mathcal{L}_{\xi}g)$ vanishes when $\xi$ is a Killing vector.
In the above calculation, we found that the horizon shift vector $\hat{\beta}$ is not a Killing vector at $v=0$ (unless $\beta(x)$ does not depend on $x$). Nevertheless, the pre-symplectic form  $\mathbf{\Omega}^{(G)}(g,\delta g, \mathcal{L}_{\hat{\beta}}g)$ vanishes at $v=0$ as we shall see in section~\ref{sec:thermo_symplectic}. 

\section{Microscopic thermodynamics on horizon}
\label{sec:thermo}

In this section, we will derive microscopic thermodynamic relations which are localized on the horizon.

\subsection{Noether charge from shifted vector}

\begin{figure}
\centering
\includegraphics[width=0.45\textwidth]{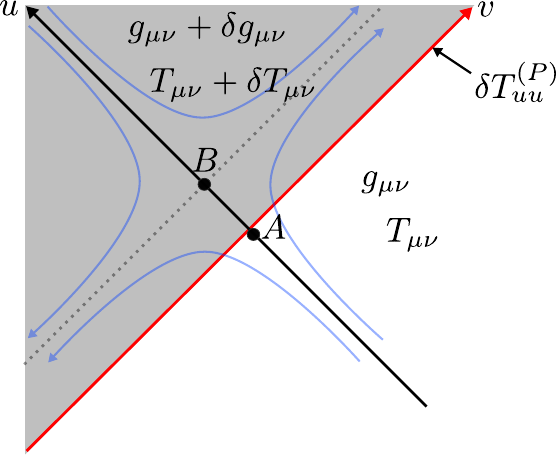}
\caption{The Noether charge integral. A redline represents the gravitational shock wave running at $u=0$. Blue lines represent the null-shifted timelike vector. The shaded region ($u>0$) experiences the metric variations $\delta g_{\mu\nu}$ and the energy momentum variations $\delta T_{\mu\nu}$ which result from the shift $v \rightarrow v + \theta(u)\alpha(x)$. In addition, at $u=0$, the variation $\delta T_{uu}^{(P)}$ from the shock wave is introduced. 
}
\label{fig-variation}
\end{figure}

Let us briefly recall the setup. The original static spacetime was given by Eq.~\eqref{eq:original} which is reprinted below
\begin{equation}
\begin{split}
ds^2 &= 2 F(u,v) dudv + G(u,v) h_{ij} dx^i dx^j \\ 
 T &= 2T_{uv}dudv + T_{uu}dudu + T_{vv}dvdv  + T_{ij}dx^{i}dx^{j}.
\end{split}
\end{equation}
The shifted metric and the energy momentum tensor are given by Eq.~\eqref{eq:shockwave} and Eq.~\eqref{eq:shockwave2} which are obtained by the substitution $v \rightarrow \tilde{v}=v + \theta(u)\alpha(x)$. 
It is worth noting that, at the leading order in $\alpha(x)$, the metric variation is given by
\begin{align}
\delta g_{\mu\nu} = \theta(u)\mathcal{L}_{\hat{\alpha}}g_{\mu\nu}, \qquad \hat{\alpha} \equiv \alpha(x)\partial_v
\end{align}
which can be verified from results in section~\ref{sec:Killing}.
In addition, there is an extra source of the energy momentum tensor $T^{(P)}_{uu}$ at $u=0$ as in Eq.~\eqref{eq:point-particle}. See Fig.~\ref{fig-variation} for summary.

Let us specify how the integral of the Noether charge is taken. We will consider an integral of the Noether charge relation Eq.~\eqref{eq:Noether} over the $(d-1)$-dimensional surface $\Lambda$ at $v=0$ that connects two $(d-2)$-dimensional boundaries $\Sigma_A$ and $\Sigma_B$ (see Fig.~\ref{fig-Killing}(b) or Fig.~\ref{fig-variation}):
\begin{align}
\Sigma_A = \{ (u,v)= (0_- , 0) \} \qquad \Sigma_B = \{ (u,v)= (\beta(x) , 0) \}.
\end{align}
Here $0_-$ means that $u$ approaches zero from below. For our purpose of deriving the shock wave equations of motion, it actually suffices to choose $\Sigma_{A}$ as any boundary with $u<0$. 

Finally, let us construct the Noether charge. As a vector field, we shall take the shifted timelike vector: 
\begin{align}
\chi \equiv \xi^{(t)} + \kappa \hat{\beta} = \kappa\Big( \big(\beta(x) - u  \big)  \partial_u + v \partial_v \Big) 
\end{align}
in order to construct the Noether charge. It is essential to observe
\begin{align}
\chi = 0 \qquad \text{at $\Sigma_{B}$}
\end{align}
which will be useful in simplifying the thermodynamic relation.

\subsection{Thermodynamic relation}

Our central formula is Eq.~\eqref{eq:Noether} which is reprinted below:
\begin{align}
\delta d \mathbf{Q}_{\chi, \text{field}} =  \mathbf{\Omega}^{(G)}(\delta g, \mathcal{L}_{\chi}g) + d\big( \chi \cdot \mathbf{\Theta}^{(G)}(\delta g) \big) + \delta \mathbf{J}^{(M)}_{\chi, \text{field}}+\frac{1}{2}(\chi\cdot\pmb{ \epsilon}) T^{\mu\nu} \delta g_{\mu\nu} \quad \text{$\{\delta g, \delta \phi\}$ on shell}. 
\end{align}
Integrating it on the $v=0$ surface generates
\begin{align}
\delta  \int_{\Lambda} d \mathbf{Q}_{\chi, \text{field}} =  \int_{\Lambda} \mathbf{\Omega}^{(G)}(\delta g, \mathcal{L}_{\chi}g) + \int_{\Lambda} d \big(\chi \cdot \mathbf{\Theta}^{(G)}(\delta g) \big) + \delta\int_{\Lambda}  d \mathbf{Q}^{(M)}_{\chi, \text{field}}. 
\end{align}

Two comments follow. First, the last term $\frac{1}{2}(\chi\cdot\pmb{ \epsilon}) T^{\mu\nu} \delta g_{\mu\nu}$ vanishes since the integral is along the vector field $\chi$ and thus $\int_{\Lambda}\ (\chi\cdot\pmb{\epsilon})=0$. Second, we have introduced the matter charge $\mathbf{Q}^{(M)}_{\chi, \text{field}}$. Here it is useful to note that, while $\hat{\beta}$ is not a Killing vector of the spacetime, one can still construct the matter charge on the $v=0$ surface since 
\begin{align}
\nabla_{\mu} ({J_{\chi,\text{field}}}^{\mu}) = - \frac{1}{2}T^{\mu\nu}(\mathcal{L}_{\chi} g_{\mu\nu}) = - T^{vi} (\mathcal{L}_{\chi} g_{vi}) = 0 \qquad (v = 0).
\end{align}

Then, by using the Stokes' theorem, we obtain
\begin{equation}
\begin{split}
\int_{\Sigma_{A}}\left( \delta  \mathbf{Q}_{\chi, \text{field}}-\delta \mathbf{Q}^{(M)}_{\chi, \text{field}}-\chi \cdot \mathbf{\Theta}^{(G)}(\delta g) \right)+ \int_{\Lambda} \mathbf{\Omega}^{(G)}(\delta g, \mathcal{L}_{\chi}g)  \\  = \int_{\Sigma_{B}} \left(\delta  \mathbf{Q}_{\chi, \text{field}}-\delta \mathbf{Q}^{(M)}_{\chi, \text{field}}-\chi \cdot \mathbf{\Theta}^{(G)}(\delta g)\right). \label{eq:thermodynamics-shock}
\end{split}
\end{equation}
Below, we will further simplify this expression by using the properties of the shifted vector field $\chi$.

Let us evaluate each term in Eq.~\eqref{eq:thermodynamics-shock} at $\Sigma_A$ and $\Sigma_{B}$. First, we observe 
\begin{align}
\chi \cdot \mathbf{\Theta}^{(G)}(\delta g) =0  \qquad \text{at $\Sigma_A,\Sigma_B$}
\end{align}
by noting that $\chi=0$ at $\Sigma_B$ and $\delta g =0$ at $\Sigma_A$. Second, we observe 
\begin{align}
\delta \mathbf{Q}_{\chi, \text{field}} = 0 \qquad \text{at $\Sigma_A$}
\end{align}
since $\delta g =0$ at $\Sigma_A$. Third, we can construct the matter charge $\mathbf{Q}^{(M)}_{\chi, \text{field}}$ as an integral over $v=0$ surface in the $u$ direction 
:
\begin{equation}
\begin{split}
\mathbf{Q}^{(M)}_{\chi, \text{field}} &\equiv \int^{u}_{-\infty} \mathbf{J}^{(M)}_{\chi, \text{field}} = - \int^{u}_{-\infty} T^{\mu\nu}\chi_{\nu}\sqrt{-g} (\mathbf{d^{d-1}x})_\mu 
= - \int^{u}_{-\infty} T^{vv}\chi_{v} \sqrt{-g}(\mathbf{d^{d-1}x})_v.
\end{split}
\end{equation}

Due to the constraint $T^{vv}=0$ at $v=0$ in Eq.~\eqref{eq:v-assumption}, we have
\begin{equation}
\delta \mathbf{Q}^{(M)}_{\chi, \text{field}} = - \int^{u}_{-\infty} (\delta T^{vv})\chi_{v} \sqrt{-g}(\mathbf{d^{d-1}x})_v.
\end{equation}
There are two potential contributions to the variation $\delta T^{vv}$. The first contribution is due to the metric shift which is explicitly given by
\begin{align}
\delta T^{vv} = T^{vv}(u,v+\alpha) - T^{vv}(u,v) \qquad u>0.
\end{align}
It turns out that this contribution is subleading in $\alpha$. Recall that the timelike vector $\xi^{(t)}$ is the symmetry of the spacetime. We then have
\begin{align}
\mathcal{L}_{\xi^{(t)}} T^{vv} = - u \partial_u T^{vv} + v \partial_v T^{vv} - 2 T^{vv} = 0.
\end{align}
Let us Taylor expand $T^{vv}$ for small $u$ and $v$. Then we find that only the terms of the form $u^a v^{a+2}$ are allowed. This suggests that $T^{vv}$ is $\sim v^2$ at most, and hence $\delta T^{vv}$ is $\sim v\alpha + \alpha^2$. Integrating it along $v=0$ only generates a contribution of $O(\alpha^2)$. The second contribution comes from the shock wave. Considering $\delta T^{vv}=4p^{(v)}\delta (u)\delta(x)$ at $v=0$, we find 
\begin{align}
\delta \mathbf{Q}^{(M)}_{\chi, \text{field}} = 0 \qquad \text{at $\Sigma_A$}
\end{align}
and 
\begin{equation}
\begin{split}
\delta \mathbf{Q}^{(M)}_{\chi, \text{field}} &= - \int^{u=\beta(x)}_{-\infty} 4p^{(v)}F\delta(u)\delta(x)\kappa(\beta(x)-u)\sqrt{-g}(\mathbf{d^{d-1}x})_v\\
&=  (\mathbf{d^{d-2}x})_{uv} \sqrt{\mathcal{G}} 4 p^{(v)}F\delta(x) \kappa \beta(x)  \qquad \text{at $\Sigma_B$}. \label{eq:Matter_Charge}
\end{split}
\end{equation}

Note that  $\sqrt{\mathcal{G}}$ is the volume element induced on the $(d-2)$-dimensional surface $(u,v)=(0,0)$. Hence, we obtain 
\begin{align}
\boxed{ \ 
 \delta\int_{\Sigma_{B}}\left(  \mathbf{Q}_{\chi, \text{field}}- \mathbf{Q}^{(M)}_{\chi, \text{field}}\right) - \int_{u=0_{-}}^{u=\beta(x)} \mathbf{\Omega}^{(G)}(\delta g, \mathcal{L}_{\chi}g)   = 0. 
\ }
\end{align}

\subsection{Pre-symplectic form}\label{sec:thermo_symplectic}

The remaining task is to evaluate the pre-symplectic form $\mathbf{\Omega}^{(G)}(\delta g, \mathcal{L}_{\chi}g)$. 
Here we will show that 
\begin{align}
\int_{u=0_{-}}^{u=\beta(x)} \mathbf{\Omega}^{(G)}(\delta g, \mathcal{L}_{\chi}g)=0.
\end{align}

Since the timelike vector $\xi^{(t)}$ is the symmetry of the original spacetime, we have \linebreak $\mathbf{\Omega}^{(G)}(\delta g, \mathcal{L}_{\xi^{(t)}}g) \simeq 0$, and
\begin{align}
\mathbf{\Omega}^{(G)}(\delta g, \mathcal{L}_{\chi}g) \simeq\kappa\mathbf{\Omega}^{(G)}(\delta g, \mathcal{L}_{\hat{\beta}}g)
\end{align}
where $\simeq$ denotes that we evaluate the differential form on the $v=0$ surface. Recalling $\delta g =  \theta(u) \mathcal{L}_{\hat{\alpha}}g$ at the linear order in $\alpha(x)$, we arrive at 
\begin{align}
\int_{u=0_{-}}^{u=\beta(x)}\mathbf{\Omega}^{(G)}(\delta g, \mathcal{L}_{\chi}g) = \kappa \int_{u=0}^{u=\beta(x)} \mathbf{\Omega}^{(G)}(\mathcal{L}_{\hat{\alpha}}g, \mathcal{L}_{\hat{\beta}}g). 
\end{align}

Let us recall that the variation of the gravity Lagrangian is given by $\delta \mathbf{L}^{(G)} = {\mathbf{E}^{(G)}}^{\mu\nu} \delta g_{\mu\nu} + d\mathbf{\Theta}(\delta g)$. Using the equations of motion and taking $\delta = \mathcal{L}_{\hat{\beta}}$, we obtain 
\begin{align}
\mathcal{L}_{\hat{\beta}} \mathbf{L}^{(G)} = - \frac{1}{2}\pmb{\epsilon}T^{\mu\nu}\mathcal{L}_{\hat{\beta}} g_{\mu\nu} + d\mathbf{\Theta}_{\hat{\beta}}. 
\end{align}
Taking another variation $\mathcal{L}_{\hat{\alpha}}$, integrating it and using the Cartan's formula, we have 
\begin{equation}
\begin{split}
\int \mathcal{L}_{\hat{\alpha}} \mathcal{L}_{\hat{\beta}} \mathbf{L}^{(G)} 
&= - \frac{1}{2} \int \mathcal{L}_{\hat{\alpha}}(\pmb{\epsilon}T^{\mu\nu}\mathcal{L}_{\hat{\beta}} g_{\mu\nu}) + \int d\mathcal{L}_{\hat{\alpha}}\mathbf{\Theta}_{\hat{\beta}}\\
&= - \frac{1}{2}\int d (\hat{\alpha}\cdot \pmb{\epsilon}T^{\mu\nu}\mathcal{L}_{\hat{\beta}} g_{\mu\nu}) + \int d\mathcal{L}_{\hat{\alpha}}\mathbf{\Theta}_{\hat{\beta}} \\ 
&= - \frac{1}{2}(\hat{\alpha}\cdot \pmb{\epsilon}T^{\mu\nu}\mathcal{L}_{\hat{\beta}} g_{\mu\nu} ) + \mathcal{L}_{\hat{\alpha}}\mathbf{\Theta}_{\hat{\beta}}\\
&=- \frac{1}{2} \alpha(x) \sqrt{-g} T^{\mu\nu}\mathcal{L}_{\hat{\beta}} g_{\mu\nu} (dx)_{v} + \mathcal{L}_{\hat{\alpha}}\mathbf{\Theta}_{\hat{\beta}} \\
&\simeq - \alpha(x) \sqrt{-g} T^{vi}\mathcal{L}_{\hat{\beta}} g_{vi} (dx)_{v} + \mathcal{L}_{\hat{\alpha}}\mathbf{\Theta}_{\hat{\beta}} 
= \mathcal{L}_{\hat{\alpha}}\mathbf{\Theta}_{\hat{\beta}}.
\end{split}
\end{equation}
Similarly, we have 
\begin{align}
\int \mathcal{L}_{\hat{\beta}}\mathcal{L}_{\hat{\alpha}}  \mathbf{L}^{(G)} = - \frac{1}{2} \beta(x) \sqrt{-g} T^{\mu\nu}\mathcal{L}_{\hat{\alpha}} g_{\mu\nu} (dx)_{u} + \mathcal{L}_{\hat{\beta}}\mathbf{\Theta}_{\hat{\alpha}}.
\end{align}
Since $\mathcal{L}_{\hat{\alpha}}$ and $\mathcal{L}_{\hat{\beta}}$ commute, we have $\int \mathcal{L}_{\hat{\beta}}\mathcal{L}_{\hat{\alpha}}  \mathbf{L}^{(G)} = \int \mathcal{L}_{\hat{\alpha}} \mathcal{L}_{\hat{\beta}} \mathbf{L}^{(G)}$. Hence we obtain~
\footnote{
For general relativity, we have checked that $\mathcal{L}_{\hat{\alpha}}\mathbf{\Theta}_{\hat{\beta}} \simeq \mathcal{L}_{\hat{\beta}}\mathbf{\Theta}_{\hat{\alpha}} \simeq 0$ via brute force calculation.
}
\begin{align}
 \mathbf{\Omega}^{(G)}(g,\mathcal{L}_{\hat{\alpha}}g, \mathcal{L}_{\hat{\beta}}g) &\equiv  \mathcal{L}_{\hat{\alpha}}\mathbf{\Theta}_{\hat{\beta}} - \mathcal{L}_{\hat{\beta}}\mathbf{\Theta}_{\hat{\alpha}}\\ 
 &\simeq -\frac{1}{2} \beta(x) \sqrt{-g} T^{\mu\nu}\mathcal{L}_{\hat{\alpha}} g_{\mu\nu} (dx)_{u}.
\end{align}
By integrating $\mathbf{\Omega}^{(G)}(g,\mathcal{L}_{\hat{\alpha}}g, \mathcal{L}_{\hat{\beta}}g) $ on the $v=0$ surface, we obtain 
\begin{align}
\int_{u=0}^{u=\beta(x)} \mathbf{\Omega}^{(G)}(\mathcal{L}_{\hat{\alpha}}g, \mathcal{L}_{\hat{\beta}}g) = 0
\end{align}
since the term with $(dx)_{u}$ does not contribute to the integral. 

This enables us to obtain the following simple constraint that governs microscopic thermodynamics of gravitational shock waves on the horizon:
\begin{align}
\boxed{ \
\delta \int_{\Sigma_B} \mathbf{Q}_{\chi, \text{field}}  - \delta  \int_{\Sigma_B}  \mathbf{Q}^{(M)}_{\chi, \text{field}} =0. \label{eq:constraint}  \
}
\end{align}
This is the central result of this paper. 

As we will see in the next section, the first term corresponds to the area deformation under two shifts $u \rightarrow u + \beta(x)$ and $v \rightarrow v + \alpha(x)$. The second term corresponds to the matter charge variation due to the gravitational shock wave, probed by the horizon translation vector field $\hat{\beta}$. It is worth emphasizing that, if the original unshifted timelike vector $\xi^{(t)}$ were used, we would not have any interesting thermodynamic relation since $\delta \mathbf{Q}^{(M)}_{\xi, \text{field}}=\delta \mathbf{Q}_{\xi, \text{field}}=0$. This is related to the fact that the area variation by $v\rightarrow v + \alpha(x)$ vanishes at the bifurcate surface as we shall explicitly see in the next section.

\section{Shock wave from area minimization}\label{sec:area}

In this section, we derive the gravitational shock wave equations of motion by using the microscopic thermodynamic relation on the horizon.

\subsection{Area variation}

We have evaluated the matter charge $\delta \mathbf{Q}^{(M)}_{\chi, \text{field}} $ in the previous section. Here we evaluate $\delta \mathbf{Q}_{\chi, \text{field}} $ at $\Sigma_B$. For general relativity with scalar fields, the Noether charge is given by~\footnote{For scalar fields, we have $\mathbf{Q}_{\chi, \text{field}}= \mathbf{Q}_{\chi, \text{Wald}}$. The difference between $\mathbf{Q}_{\chi, \text{field}}$ and $\mathbf{Q}_{\chi, \text{Wald}}$ generically vanishes as $\chi\rightarrow 0$. See section~\ref{sec:refinement} for discussions.}
\begin{align}
{\mathbf{Q}_{\chi, \text{field}}}_{a}  = - \frac{1}{16\pi} \pmb{\epsilon}_{abc} \nabla^b \chi^c
\end{align}
where $a=a_1\cdots a_{d-2}$ collectively denotes $(d-2)$ indices. Recall that $\chi$ is held fixed in variations. Also $\chi=0$ at $\Sigma_B$. Hence we have 
\begin{align}
\delta {\mathbf{Q}_{\chi, \text{field}}}_{a}  = - \frac{1}{16\pi} (\delta{\pmb{\epsilon}_{abc}}) \nabla^b \chi^c \qquad \text{at $\Sigma_B$}.
\end{align}
Non-trivial contributions come from $b,c=u,v$. We thus have
\begin{align}
\delta{\mathbf{Q}_{\chi, \text{field}}} = - \frac{\kappa}{8\pi} \frac{1}{F} (\mathbf{d^{d-2}x})_{uv} \delta{\sqrt{\mathcal{G}}} \qquad \text{at $\Sigma_B$}
\end{align}
where ($\mathbf{d^{d-2}x})_{uv} \sqrt{\mathcal{G}}$ is the volume element induced on the $(d-2)$-dimensional surface $\Sigma_B$. 

\begin{figure}
\centering
\includegraphics[width=0.35\textwidth]{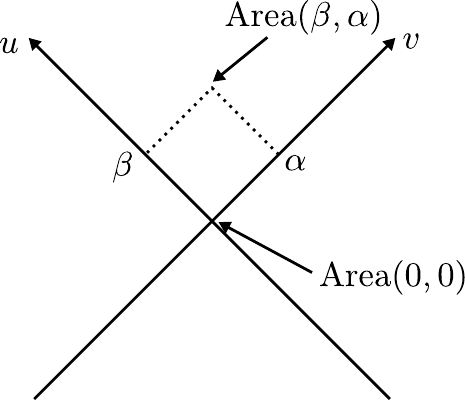}
\caption{The area deformation in the original unperturbed geometry.}
\label{fig-area}
\end{figure}

It is worth relating the variation of the Noether charge explicitly to the area deformation (Fig.~\ref{fig-area}). Let $\text{Area}(\beta,\alpha)$ denote the total area of the hypersphere at $(u,v)=(\beta,\alpha)$ in the original unperturbed metric $ds^2$. The Noether charge integral gives 
\begin{equation}
\begin{split}
\int_{\Sigma_B} \delta{\mathbf{Q}_{\chi, \text{field}}} &= 
- \frac{\kappa}{8\pi} \frac{1}{F}\Big( \text{Area}(\beta,\alpha)  - \text{Area}(\beta,0)\Big)\\
&= - \frac{\kappa}{8\pi} \frac{1}{F}\Big( \text{Area}(\beta,\alpha) -\text{Area}(0,0)\Big)
\end{split}
\end{equation}
where we have used $\text{Area}(\beta, 0) = \text{Area}(0, 0)$, since $(u,v)=(\beta,0)$ is still on the horizon. Hence the Noether charge variation is proportional to the increase of the area of the bifurcating horizon by two null-shifts $u\rightarrow u+\beta$ and $v \rightarrow v+ \alpha$. Note that $\alpha$ comes from the shift by the gravitational shock wave while $\beta$ is set by the shift in the vector field $\chi$. 

One comment follows. If we used the timelike Killing vector $\xi$ instead of the shifted one $\chi$, we would have the contribution proportional to $\text{Area}(0, \alpha) - \text{Area}(0, 0)$ which vanishes since $(u,v)=(0,\alpha)$ is still on the horizon. Hence the Noether charge relation would be trivial.

\subsection{Induced metric}

The remaining task is to evaluate the induced metric on $\Sigma_B$. A spacetime point on $\Sigma_B$ can be expressed as 
\begin{align}
X^{\mu}(\vec{y}) \equiv ( \beta(\vec{y}), \alpha(\vec{y}), \vec{y} )
\end{align}
where $\alpha(\vec{y})$ results from the metric variation. The induced metric on $\Sigma_B$ is given by 
\begin{align}
d \sigma^2 \equiv ds^2|_{(u,v,x) = X(\vec{y})} \equiv \mathcal{G}_{ij}(\vec{y}) dy^{i}dy^{j}
\end{align}
where
\begin{align}
\mathcal{G}_{ij}(\vec{y}) = g_{\mu\nu}(X(\vec{y})) \frac{\partial X^{\mu}}{\partial y^{i}}\frac{\partial X^{\nu}}{\partial y^{j}}.
\end{align}

We shall Taylor expand $\mathcal{G}_{ij}(\vec{y})$ for small $\alpha,\beta$. Looking at $(\mu,\nu)=(u,v)$, we have the following contribution to $\mathcal{G}_{ij}(\vec{y})$:
\begin{align}
g_{uv}(X(\vec{y})) \frac{\partial \beta}{\partial y^{i}}\frac{\partial \alpha}{\partial y^{j}}  = \left( F(0,0) + \frac{\partial^2 F(u,v)}{\partial u \partial v} \alpha \beta + \cdots \right) \frac{\partial \beta}{\partial y^{i}}\frac{\partial \alpha}{\partial y^{j}}. \label{eq:sub}
\end{align}
Looking at $(\mu,\nu)=(i,j)$, we have the following contribution to $\mathcal{G}_{ij}(\vec{y})$:
\begin{align}
g_{ij}(X(\vec{y})) dy^{i}dy^{j} = \left(G(0,0) +\frac{\partial^2 G(u,v)}{\partial u \partial v} \alpha\beta  + \cdots \right)h_{ij}(\vec{y})dy^{i}dy^{j} . 
\end{align}

We observe that the leading correction to $\mathcal{G}_{ij}(\vec{y})$ is at the order of $\alpha \beta$. In the expansion of Eq.~\eqref{eq:sub}, the term $\frac{\partial^2 F(u,v)}{\partial u \partial v}\alpha\beta\frac{\partial \beta}{\partial y^{i}}\frac{\partial \alpha}{\partial y^{j}}$ is at the order of $\alpha^2 \beta^2$, and hence is negligible. Then we find 
\begin{equation}
\begin{split}
d\sigma^2 &\approx 
 2F(0,0) \frac{\partial \beta}{\partial y^{i}}\frac{\partial \alpha}{\partial y^{j}} dy^{i}dy^{j} + \left( G(0,0)  + \frac{\partial^2 G(u,v)}{\partial u \partial v} \alpha \beta \right) h_{ij}(\vec{y})  dy^{i}dy^{j} \\ 
&= d\sigma^2_{*} + 2F(0,0)  \frac{\partial \beta}{\partial y^{i}}\frac{\partial \alpha}{\partial y^{j}} dy^{i}dy^{j} + \frac{\partial^2 G(u,v)}{\partial u \partial v} \alpha \beta  h_{ij}(\vec{y})  dy^{i}dy^{j}
\end{split}
\end{equation}
where $d\sigma^2_{*}$ represents the unperturbed induced metric when $(\alpha,\beta)=(0,0)$:
\begin{align}
d\sigma_{*}^2 = G(0,0) h_{ij}dy^{i}dy^{j}.
\end{align}
To summarize, we have obtained the variation of the induced metric:
\begin{align}
\delta \mathcal{G}_{ij} =  2  F \partial_i \alpha \partial_j \beta + G_{,uv} \alpha\beta  h_{ij} \label{eq:induced-variation}
\end{align}
where $F$ and $G_{,uv}$ are evaluated at $u=v=0$. 

Recall that the first order variation of the determinant is
\begin{align}
\delta \sqrt{\mathcal{G}} = \frac{1}{2}\sqrt{\mathcal{G}}\mathcal{G}^{ij} \delta \mathcal{G}_{ij}.
\end{align}
Hence the variation of the determinant at $\Sigma_B$ is given by
\begin{align}
\boxed{ \
\delta \sqrt{ \mathcal{G} } = \sqrt{ \mathcal{G} }  \frac{1}{2G}  \Big( 2F (\partial^j \alpha) (\partial_j \beta) + (d-2) G_{,uv}\alpha \beta \Big).
\ }
\end{align}
Note $\mathcal{G}^{ij}=g^{ij}=\frac{h^{ij}}{F}$ at $u=v=0$. Observe $\partial_j \Big(\beta (\partial^j \alpha) \sqrt{h} \Big) = (\partial_j \beta) (\partial^j \alpha) \sqrt{h} + \beta \partial_j \Big( (\partial^j \alpha) \sqrt{h}\Big)$.
After integrating by parts, we obtain 
\begin{align}
\boxed{\
\delta \sqrt{ \mathcal{G} } = \sqrt{ \mathcal{G} }  \frac{1}{2G}  \Big( - 2F \beta (\Delta_{h_{ij}} \alpha) + (d-2) G_{,uv}\alpha \beta \Big). \label{eq:g-variation}
\
}
\end{align}
Note that total derivatives do not contribute to the surface integral of the Noether charge.

Finally, we derive the shock wave equations of motion. Using the expression of deformed area density Eq.~\eqref{eq:g-variation}, we obtain
\begin{align}
\delta{\mathbf{Q}_{\chi,\text{field}}} =
 -(\mathbf{d^{d-2}x})_{uv}  \frac{\kappa\sqrt{\mathcal{G}}}{8\pi}\frac{1}{2GF} \Big( - 2F \beta(\Delta_{h_{ij}} \alpha) + (d-2) G_{,uv}\alpha \beta \Big) 
\end{align} 
then the constraint Eq.~\eqref{eq:constraint} leads to
\begin{align}
 -\sqrt{\mathcal{G}} \frac{\kappa}{8\pi} \frac{1}{2GF}  \Big( - 2F (\Delta_{h_{ij}} \alpha) + (d-2) G_{,uv}\alpha  \Big) \beta =  \sqrt{\mathcal{G}} 4 p^{(v)}F\delta(x) \kappa \beta. 
\end{align}
We can arrange it into the following form
\begin{align}
  \Big(   (\Delta_{h_{ij}}\alpha) - \frac{(d-2)}{2} \frac{G_{,uv}}{F}\alpha  - 32\pi GF p^{(v)}\delta(x) \Big) \beta \kappa \sqrt{\mathcal{G}} = 0. 
\end{align}
Hence we recover the shock wave equation of motion 
\begin{align}
\Delta_{h_{ij}} \alpha(x) - \frac{d-2}{2} \frac{G_{,uv}}{F}  \alpha(x) = 32 \pi p^{(v)} G F \delta(x).
\end{align}

\section{Gauss-Bonnet gravity}
\label{sec:application}

The microscopic thermodynamic relation can be generalized to arbitrary covariant theory of gravity. The gravitational shock wave equations in the Gauss-Bonnet gravity have been previously studied for a certain black hole solution in \cite{Roberts:2014isa} while exhaustive studies have been presented in \cite{Huang:2017ohr, Huang:2018snb} with rather involved calculations. In this section, we present a simpler derivation of the gravitational shock wave equations of motion in the Gauss-Bonnet gravity based on the Noether charge method. 

\subsection{Noether charge in Gauss-Bonnet gravity}

The Lagrangian $d$-form of the Gauss-Bonnet gravity is 
\begin{align}
\mathbf{L}= \pmb{\epsilon}_{a_1\cdots a_d} \Big( \frac{1}{16\pi} R + \frac{\lambda}{16\pi} ( R_{abcd}R^{abcd} - 4 R_{ab}R^{ab} + R^2  )  \Big).
\end{align}
This yields the Noether charge $(d-2)$-form~\cite{Iyer:1994ys}
\begin{align}
{\mathbf{Q}_{\chi}}_{a} = - \pmb{\epsilon}_{ade} \Big( \frac{1}{16\pi} \nabla^d \chi^e + \frac{\lambda}{8\pi} (R \nabla^d \chi^e + 4 \nabla^{[f} \chi^{d]} R^{e}_{\; \; f} + R^{defh} \nabla_f \chi_h ) \Big). \label{eq:GB-charge}
\end{align}

We shall focus on a family of the metric of Eq.~\eqref{eq:original}. The main task is to evaluate the variation $\delta{\mathbf{Q}_{\chi}}$ under $(u,v)=(\beta,0)\rightarrow (\beta,\alpha)$ at $\Sigma_B$. It suffices to compute the contribution to ${\mathbf{Q}_{\chi}}_{a}$ from $(d,e)=(u,v)$ in Eq.~\eqref{eq:GB-charge}\footnote{Note there is a factor of 2 which comes from the contribution of $(d,e)=(v,u)$}:
\begin{align}
- (\mathbf{d^{d-2}x})_{uv}\sqrt{\mathcal{G}} \frac{1}{8\pi}\left( 1 + 2 \lambda  H   \right) \nabla^{[u}\chi^{v]}, \qquad H \equiv R_{\mu\nu} g^{\mu\nu} - 4R_{uv}g^{uv} - 2 R^{u \; \;  \; \; }_{\; \; u v u} g^{uv}
\end{align}
and its variation $\delta{\mathbf{Q}_{\chi}}_{a}$:
\begin{align}
- (\mathbf{d^{d-2}x})_{uv}(\delta\sqrt{\mathcal{G}}  )\frac{1}{8\pi}\left( 1 + 2 \lambda  H   \right) \nabla^{[u}\chi^{v]}  
- (\mathbf{d^{d-2}x})_{uv}\sqrt{\mathcal{G}}  \frac{1}{8\pi} 2 \lambda  (\delta H)   \nabla^{[u}\chi^{v]} . \label{eq:GB-variation}
\end{align} 

To evaluate $H$, we need the following non-vanishing Ricci tensors:
\begin{equation}
R_{uv} = \left(  - \frac{F_{,uv}}{F}   -  \frac{d-2}{2} \frac{G_{,uv} }{G} \right) \qquad
R_{ij} = R_{ij}^{(d-2)} - \left( \frac{G_{,uv}}{F} \right)h_{ij}
\end{equation} 
and the following Riemann tensor:
\begin{align}
R^{u \; \;  \; \; }_{\; \; u v u} = \partial_{v} \Gamma^{u}_{uu} - \partial_u \Gamma^{u}_{vu} 
+ \Gamma^{u}_{v\alpha} \Gamma^{\alpha}_{uu} - \Gamma^{u}_{u\alpha} \Gamma^{\alpha}_{vu} 
= \partial_v \frac{F_{,u}}{F} \simeq \frac{F_{,uv}}{F}.
\end{align}
Putting all these together, we obtain the following simple result
\begin{align}
\boxed{\
H = R_{ij}^{(d-2)} g^{ij} = \frac{R^{(d-2)}}{G}
\
}
\end{align}
where $R^{(d-2)}$ is the scalar curvature computed from $h_{ij}$.

As for $\delta H$, the following observation simplifies the calculation. Let us focus on the variation of the first term $R_{\mu\nu} g^{\mu\nu}$ in $H$:
\begin{align}
\delta( R_{\mu\nu} g^{\mu\nu} ) = \delta( R_{\mu\nu}) g^{\mu\nu} + R_{\mu\nu} \delta( g^{\mu\nu} ).
\end{align}
By utilizing the following relation
\begin{align}
\delta R^{\rho \; \;  \; \; }_{\; \; \mu \lambda \nu} = \nabla_{\lambda} \delta \Gamma^{\rho}_{\nu\mu} - \nabla_{\nu} \delta \Gamma^{\rho}_{\lambda\mu}.
\end{align}
$\delta( R_{\mu\nu}) g^{\mu\nu}$ can be expressed as a total derivative. In a similar manner, we can show the following
\begin{align}
\delta H = R_{\mu\nu} \delta g^{\mu\nu} - 4R_{uv} \delta g^{uv} -  2 R^{u \; \;  \; \; }_{\; \; u v u} \delta g^{uv} + \text{(total derivative)}.
\end{align}
Note that total derivatives do not contribute to the surface integral of the Noether charge. By dropping these terms, we obtain
\begin{align}
\delta H = R_{ij}^{(d-2)} \delta \mathcal{G}^{ij}. \
\end{align}
Using $\delta \sqrt{\mathcal{G}}= \frac{1}{2}\sqrt{\mathcal{G}} \mathcal{G}^{ij}\delta \mathcal{G}_{ij} = - \frac{1}{2}\sqrt{\mathcal{G}} \mathcal{G}_{ij}\delta \mathcal{G}^{ij}$, the total variation Eq.~\eqref{eq:GB-variation} can be expressed as 
\begin{align}
 \delta{\mathbf{Q}_{\chi, \text{field}}} =
(\mathbf{d^{d-2}x})_{uv} \frac{\kappa\sqrt{\mathcal{G}}}{8\pi F}\delta \mathcal{G}^{ij} \left( \frac{1}{2}  \mathcal{G}_{ij}\Big( 1 + 2 \lambda  \frac{R^{(d-2)}}{G}   \Big) - 2 
 \lambda R_{ij}^{(d-2)}\right). \label{eq:GB-total-variation}
\end{align} 
Finally, by making use of Eq.~\eqref{eq:induced-variation} and integrating by parts, one can obtain the gravitational shock wave equations of motion in the Gauss-Bonnet gravity.

\subsection{Isotropic space}

So far, our discussions are applicable to generic static spacetime in the Gauss-Bonnet gravity. By specializing in isotropic spaces, one obtains further simplified expression of $\delta \mathbf{Q}_{\chi}$. Let us consider the following metric
\begin{align}
ds^2 = 2F(u,v)dudv + r_H^2 h_{ij} dx^i dx^j
\end{align}
where the metric of $h_{ij}$ depends on $k=1,0,-1$ and looks like
\begin{equation}
\begin{split}
&k =1 \quad \rightarrow \ \text{de Sitter (sphere)} \\
&k =0  \quad \rightarrow \ \text{flat space} \\
&k=-1  \ \rightarrow \ \text{anti-de Sitter (hyperbolic)}.
\end{split}
\end{equation}
We then have 
\begin{align}
R_{ij}^{(d-2)} = (d-3)k h_{ij} \qquad R^{(d-2)} = k (d-3)(d-2).
\end{align}
Plugging these into Eq.~\eqref{eq:GB-total-variation}, we obtain 
\begin{align}
\delta{\mathbf{Q}_{\chi, \text{field}}} =
 -(\mathbf{d^{d-2}x})_{uv}  \frac{\kappa{\sqrt{\mathcal{G}}}}{8\pi} \frac{1}{2GF}\left( 1 + 2 \lambda  \frac{k (d-3)(d-4)}{r_{H}^2}   \right) \Big( - 2F\beta (\Delta_{h_{ij}} \alpha) + (d-2) G_{,uv}\alpha \beta \Big) 
\end{align} 
where $r=r_{H}$ is the horizon radius. By setting $\lambda_{GB}\equiv \lambda \frac{k (d-3)(d-4)}{r_{H}^2}$, we see that the Noether charge variation $\delta \mathbf{Q}_{\chi, \text{field}}$ is rescaled by a factor of $(1 + 2\lambda_{GB})$. This recovers the result from \cite{Roberts:2014isa}
\begin{align}
   (1+2\lambda_{GB})\Big( \Delta_{h_{ij}} \alpha(x) - \frac{d-2}{2} \frac{G_{,uv}}{F}  \alpha(x)\Big) = 32 \pi p^{(v)} G F \delta(x).
\end{align}

\section{Measuring shock waves}\label{sec:OTOC}

In this section, we present a brief discussion on the connection between OTOCs and the thermodynamic relation. Here we shall focus on the cases where the infalling and outgoing matters consist of scalar fields for simplicity of discussion. 

So far, we have studied geometries with single gravitational shock waves running in the $v$ direction and derived the thermodynamic relations by using shifted vector fields in the $u$ direction. A naturally arising question concerns geometries with two intersecting gravitational shock waves running in the $u$ and $v$ directions. Such geometries can be generated by using shifts $\alpha(x)$ and $\beta(x)$ in the $v$ and $u$ coordinates respectively, and have been explicitly considered by Kiem, Verlinde and Verlinde~\cite{Kiem:1995iy}. Readers might wonder if similar analyses based on the covariant phase space formalism would reveal thermodynamic relations for two gravitational shock waves. Unfortunately, the horizon area deformation induced by two gravitational shock waves are the second-order contributions, and as such, naive application of the covariant phase space formalism does not appear to give rise to non-trivial thermodynamic relations~\footnote{We expect that a careful analysis, similar to the one given in this paper for higher-order matter field variations, will enable us to treat higher-order metric variations in a controlled matter.}. 

Nevertheless, the microscopic thermodynamic relation can be generalized to geometries with two shock waves in a straightforward manner. This is due to the fact that, in a geometry with two shock waves, each shock wave needs to satisfy the corresponding gravitational equation of motion, which is identical to the one with a single shock wave~\cite{Kiem:1995iy}. Then the effective Lagrangian for two shock waves can be constructed as follows:
\begin{align}
\boxed{
L_{\text{shock}} = \text{Area}(\alpha(x),\beta(x)) - \int \alpha(x) T^{(u)}(x) -  \int \beta(x) T^{(v)}(x)  \label{eq:two-shock}
}
\end{align}
up to a multiplicative factor for the area term~\footnote{The effective Lagrangian similar to Eq.~\eqref{eq:two-shock} was derived by 't Hooft~\cite{HOOFT:1996aa} for the Schwarzschild black hole.}. Here $T^{(u)}(x)$ and $T^{(v)}(x)$ are energy-momentum tensor profiles running in the $u$ and $v$ directions respectively. 
By taking the variation of the Lagrangian, we indeed recover equations of motion for $\alpha(x)$ and $\beta(x)$:
\begin{align}
\frac{ \delta L_{\text{shock}} }{\delta \alpha(x)} = E_{\text{shock}}(\beta(x), T^{(u)}(x)), \qquad 
\frac{\delta L_{\text{shock}}}{\delta \beta(x)} = E_{\text{shock}}(\alpha(x), T^{(v)}(x)).
\end{align}
Here one may interpret $\alpha(x)$ and $\beta(x)$ as scalar fields which live on the bifurcating sphere, and the $\text{Area}(\alpha(x),\beta(x))$ as interacting massive scalar fields. Then, $T^{(u)}(x)$ and $T^{(v)}(x)$ can be interpreted as source terms coupled with fields $\alpha(x)$ and $\beta(x)$. 

Let us now turn our attention to the problem of measuring gravitational shock waves. Recall that the gravitational shock waves can be probed by out-of-time order correlation (OTOC) functions of the form $\langle V(0)W(t)V(0) W(t)\rangle$ within the AdS/CFT correspondence. When the time separation between two perturbations $V(0)$ and $W(t)$ is larger than the scrambling time, two gravitational shock waves will intersect with each other near the black hole horizon. It can then be approximated that two shock waves are running on the black hole horizon and intersect with each other. Such an approximation would be valid if the values of $u,v$ at the intersecting location are smaller than the effective shifts $\alpha, \beta$.

At the heart of the calculation of OTOCs is the gravitational scattering unitary matrix. In his pioneering work, 't Hooft~\cite{HOOFT:1996aa} derived the following scattering unitary matrix for the Schwarzschild black hole (often called the horizon $S$-matrix)~\footnote{see~\cite{Polchinski:2015cea} also.}:
\begin{align}
\mathcal{U} = \exp \left[ i \iint  d \Omega_1 d \Omega_2 P_{\text{out}}(\Omega_1) f(\Omega_1,\Omega_2) P_{\text{in}}(\Omega_2)   \right]
\end{align}
where $f(\Omega_1,\Omega_2)$ is the green function for the shock wave equation~\footnote{
Quantum mechanical aspects of the scattering matrix have been further studied in recent years, see~\cite{Betzios:2016yaq, Gaddam:2020mwe, Gaddam:2020rxb} for instance. 
}. Here $P_{\text{in}}$ and $P_{\text{out}}$ account for the incoming and outgoing energy flux. One can cast the above scattering matrix in an illuminating form which makes the relation to the soft thermodynamics more explicit. Here we consider the limit where two shock waves collide near the horizon. Note that one can identify $P_{\text{out}}(\Omega_1)$ and $P_{\text{in}}(\Omega_2)$ as $T^{(u)}(x)$ and $T^{(v)}(x)$ respectively. Then, one obtains 
\begin{align}
\alpha(x) = \int d\Omega_2 f(\Omega_1, \Omega_2) P_{\text{in}}(\Omega_2), \qquad \beta(x) = \int d\Omega_1 f(\Omega_1, \Omega_2) P_{\text{out}}(\Omega_1).
\end{align}
Recalling that the equations of motion for $\alpha(x)$ and $\beta(x)$ read
\begin{align}
\delta \text{Area}(\alpha(x),\beta(x)) = \int \alpha(x)T^{(u)}(x), \qquad  \delta \text{Area}(\alpha(x),\beta(x)) = \int \beta(x) T^{(v)}(x),
\end{align}
one can rewrite the above expression as follows:
\begin{align}
\mathcal{U} = \exp \Big( i  (\text{Area}(\alpha(x),\beta(x))  -\text{Area}(0,0)  )  \Big)
\end{align}
up to a multiplicative factor for the area. Since $\text{Area}(0,0)$ is a constant, one may express the scattering unitary matrix simply as 
\begin{align}
\mathcal{U} = \exp \Big( i  \cdot \text{Area}(\alpha(x),\beta(x) ).  \Big)
\end{align}
This expression suggests that, at the leading order, the decorrelation of OTOCs is given by the area deformation~\footnote{Direct measurement of OTOCs in many-body quantum systems is challenging with inverse time-evolution, but an alternative formulation of OTOCs using entanglement does not require inverse time-evolution. Various experimental protocols for measuring OTOCs in controlled quantum simulators have been proposed~\cite{Swingle:2016aa, Yao:aa, Yunger-Halpern:2018aa, Dressel:2018aa, Yoshida:2019aa}, with concrete experimental demonstrations~\cite{Garttner:2017aa, Li:2017aa, Landsman:2019aa, Blok:2021aa}. It is still premature to envision measurement of OTOCs in gravitational systems, but at the conceptual level, the smoothness of the black hole horizon for an infalling observer is an indirect evidence of decorrelation of OTOCs, see~\cite{Beni19, Pasterski20} for details.}

It is worth emphasizing that this expression of the scattering unitary matrix can be generalized to any bifurcating horizon in generic static spacetime in arbitrary covariant theory of gravity~\footnote{While OTOCs are concepts primarily studied within the framework of the AdS/CFT correspondence, they can be generically defined as correlation functions of matter fields which were initially located away from the horizon. Hence, we speculate that our observation may be generalized to asymptotically flat spaces as well as the de-Sitter horizon.}. The above expression of the scattering matrix can be viewed as a concrete method of measuring the soft charge, which is realized as the gravitational shock wave. We hope to further expand this observation in the future. 

Also note that matter contents render rich additional structures in the scattering matrix, see the table in~\cite{HOOFT:1996aa} for instance. It will be interesting to extend our analysis to the cases where the infalling and outgoing matters are not scalar fields. 

\section{Outlook}\label{sec:outlook}

In this work, we have derived a simple microscopic thermodynamic relation for the gravitational shock waves by providing a generic framework to construct the Noether charge for them. 

We anticipate that our formalism can be applied to a wide variety of geometries in order to study the effect of the gravitational backreaction. One interesting future problem is to apply a similar method to stationary black holes where thermodynamic treatment involves angular momentum~\cite{BenTov:2017kyf}.

We did not follow the standard treatment of the soft hair physics in characterizing the gravitational shock waves. Instead, we opted to work on the gravitational shock wave solutions on a more generic setting. It would be interesting to translate our results to the language of the soft hair physics and study the consequence of the finiteness of the black hole entropy. Also, it is an interesting future problem to generalize the expression of the gravitational scattering matrix so that it would be applicable to generic soft charges which may be beyond the gravitational shock waves. Developing a generic framework for measurement of soft charges is also an interesting problem.

\subsection*{Acknowledgment}

We would like to thank Yoni BenTov, Yiming Chen, Laura Donnay and Monica Pate for useful discussion. This work was completed partly as SL's master thesis. We would like to thank the Perimeter Scholars International for support during this research. Research at Perimeter Institute is supported by the Government of Canada through Industry Canada and by the Province of Ontario through the Ministry of Research \& Innovation. SL and BY are supported in part by Discovery Grants from the Natural Sciences and Engineering Research Council of Canada. 

\bibliographystyle{utphys}

\providecommand{\href}[2]{#2}\begingroup\raggedright\endgroup

\end{document}